\newif\ifNotes\Notesfalse
\newif\ifAnon\Anonfalse
\newif\ifDraft\Draftfalse
\newif\ifArxiv\Arxivfalse
\newif\ifCamera\Camerafalse
\newif\ifPageNum\PageNumtrue

\ifCamera
  \Notesfalse
  \Anonfalse
  \Draftfalse
  \Arxivfalse
  \PageNumfalse
\fi

\documentclass[conference,10pt,twocolumn]{IEEEtran}
\ifPageNum
\fi
\usepackage[utf8]{inputenc}
\usepackage[english]{babel}
\usepackage{blindtext}
\usepackage{enumitem}
\usepackage{authblk}
\usepackage[numbers,sort]{natbib}
\usepackage{xargs} 
\usepackage{booktabs}

\usepackage{courier}
\usepackage{amsmath}
\usepackage{MnSymbol}
\usepackage[official]{eurosym}
\usepackage[T1]{fontenc}
\usepackage{bbding}
\usepackage{blindtext}
\usepackage{graphicx}
\graphicspath{ {./images/} }
\usepackage[pdftex,dvipsnames]{xcolor} 
\usepackage{hyperref}
\hypersetup{
    colorlinks=true,
    linkcolor=blue,
    filecolor=magenta,      
    urlcolor=blue,
}
\usepackage[nameinlink,capitalise,noabbrev]{cleveref}

\definecolor{TolDarkGreen}{HTML}{117733}

\usepackage{pifont}

\usepackage{tikz}
\usetikzlibrary{arrows.meta}
\usetikzlibrary{shapes}
\usetikzlibrary{decorations.markings}
 
\newcommand*\halfcirc{\begin{tikzpicture}\draw (0,0) circle (1.0ex); \fill (0,0)-- (90:1ex) arc (90:270:1ex) -- cycle;\end{tikzpicture}}
\newcommand*\fullcirc{\tikz\fill (0,0) circle (1.0ex);} 

\newcommand{\xmark}{\leavevmode{\color{red}\ding{55}}\xspace}%
\newcommand{\cmark}{\leavevmode{\color{TolDarkGreen}\ding{51}}\xspace}%
\newcommand{\nmark}{--}

\newcommand{\swallow}[1]{}
\ifNotes
  \newcommand{\colorcomment}[2]{\leavevmode\unskip\space{\color{#1}#2}\xspace}
\else
  \newcommand{\colorcomment}[2]{\leavevmode\unskip\relax}
\fi

\ifArxiv

\else

\fi

\begin{document}
\title{SoK: Design Tools for Side-Channel-Aware Implementations}

\ifAnon
\author{}
\else
\author{Ileana Buhan\IEEEauthorrefmark{1}}
\author{Lejla Batina\IEEEauthorrefmark{1}}
\author{Yuval Yarom\IEEEauthorrefmark{2}}
\author{Patrick Schaumont\IEEEauthorrefmark{3}}
\affil{\IEEEauthorrefmark{1} Radboud University, Digital Security \\
\IEEEauthorrefmark{2} University of Adelaide and Data61 \\
\IEEEauthorrefmark{3} Worcester Polytechnic Institute
}
\fi
\date{}
\maketitle

\begin{abstract}
Side-channel attacks that leak sensitive information through a computing device's interaction with its physical environment have proven to be a severe threat to devices' security, particularly when adversaries have unfettered physical access to the device.
Traditional approaches for leakage detection measure the physical properties of the device.
Hence, they cannot be used during the design process and fail to provide root cause analysis.
An alternative approach that is gaining traction is to automate leakage detection by modeling the device.
 The demand to understand the scope, benefits, and limitations of the proposed tools intensifies with the increase in the number of proposals.

In this SoK, we classify approaches to automated leakage detection based on the model's source of truth.
We classify the existing tools on two main parameters: whether the model includes measurements from a concrete device and the abstraction level of the device specification used for constructing the model.
We survey the proposed tools to determine the current knowledge level across the domain and identify open problems.
In particular, we highlight the absence of evaluation methodologies and metrics that would compare proposals' effectiveness from across the domain.
We believe that our results help practitioners who want to use automated leakage detection and researchers interested in advancing the knowledge and improving automated leakage detection.

\end{abstract}
\section{Introduction}
When a computing device operates, it interacts with its physical environment.
In his seminal work, Kocher~\cite{Koc96} demonstrated that the power consumption of a device leaks information about the data it processes, allowing the recovery of cryptographic keys.
Since then, research has demonstrated leakage of sensitive information via other \emph{side-channels}, including electromagnetic emanations (EM)~\cite{GMO01,QS01}, timing~\cite{Bernstein05,BrumleyT11,Pag02},  micro-architectural components~\cite{GeYCH18,abs-2103-14244, BZB+05}, and even acoustic and photonic emanations~\cite{GenkinST14,KN+13}.

\begin{figure}[b]
\centering
\includegraphics[width=0.45\textwidth]{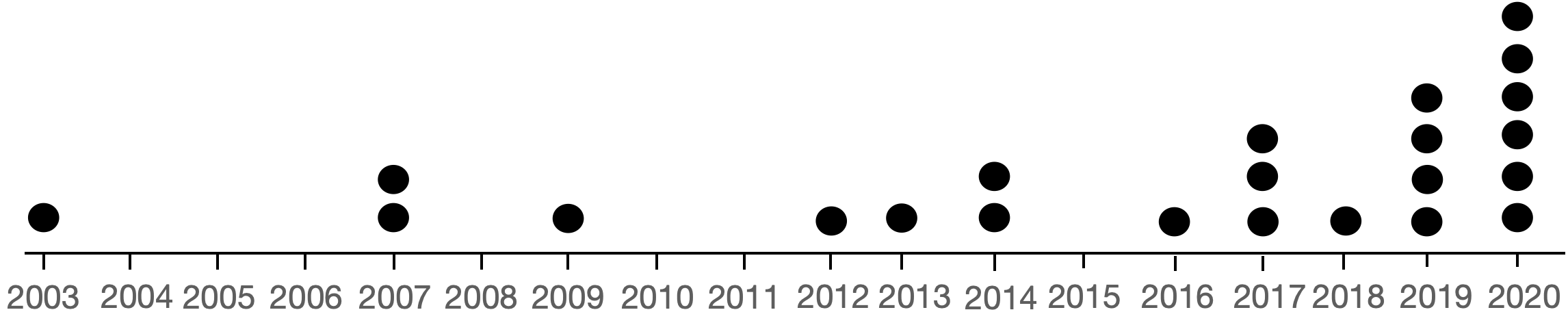}
\caption{Distribution of papers which propose a new simulator per year (for multiple publications on the same tool,  we cite the most recent one).}
\label{fig:distribution_sim}
\end{figure}

In response to developments in attacks, several methodologies for leakage detection and assessment have been developed.
Typically, such techniques emulate an attack.  
They involve collecting side-channel traces, e.g., power traces, from the device and analyzing these traces to demonstrate an attack or the existence of leaks.
While effective, such methodologies require the physical device's presence for evaluation, and this demand poses significant challenges.

In the pre-silicon stage of the development, the device does not yet exist; hence it cannot be adequately assessed.
Conversely, in the post-silicone stage, detailed design information may not be accessible, for example, when using third-party components.
Consequently, it may be challenging to identify the root cause of leakage.
Moreover, detecting, verifying, and mitigating side-channel leaks require expert knowledge and expensive equipment.

In recent years, we see the emergence of an alternative approach for evaluating device resilience to side-channel attacks.
Instead of measuring the leakage from a physical device, \emph{leakage emulators} aim to evaluate the device's model to reduce the effort required for leakage detection and potentially perform leakage detection early in the development process.
The appeal of automation is proven by the early attempts of creating such tooling and the increased recent efforts directed to this purpose, as demonstrated in \cref{fig:distribution_sim}.  
However, the abundance of proposed tools does not necessarily offer a solution for practitioners. 
Each tool aims to address a specific scenario, and with the increasing number of proposals, it may be complex to identify the best tool for each use case.
Moreover, comparison of tools across the domain is lacking, preventing a straightforward assessment of the benefits that each of the tools may offer.
A comprehensive study of automated tooling available for computer-aided cryptography was recently published~\cite{Barbosa_2021_Computer_aided_crypto} which covers 
design, functional verification, and implementation-level security of digital side-channels.  The study covers tooling for side-channels such as execution time and side effects in shared resources (e.g., cache). Still, it excludes physical side-channels such as power consumption or EM radiation. This paper covers the gap and presents a taxonomy of state-of-the-art tooling for protecting against physical side-channel attacks.
This work aims to chart the landscape and present a coherent view of current advances in leakage evaluation automation.
Specifically, it aims to
\begin{itemize}
  \item Give a comprehensive survey of the available tooling for leakage detection, verification, and mitigation and clarify the current capabilities and limitations.
  \item Present a taxonomy for the published tools, bringing forward their main innovations and potentials.
  \item Outline existing challenges and promising new research directions.
\end{itemize}

\begin{figure}[t]
\centering
\includegraphics[width=0.5\textwidth]{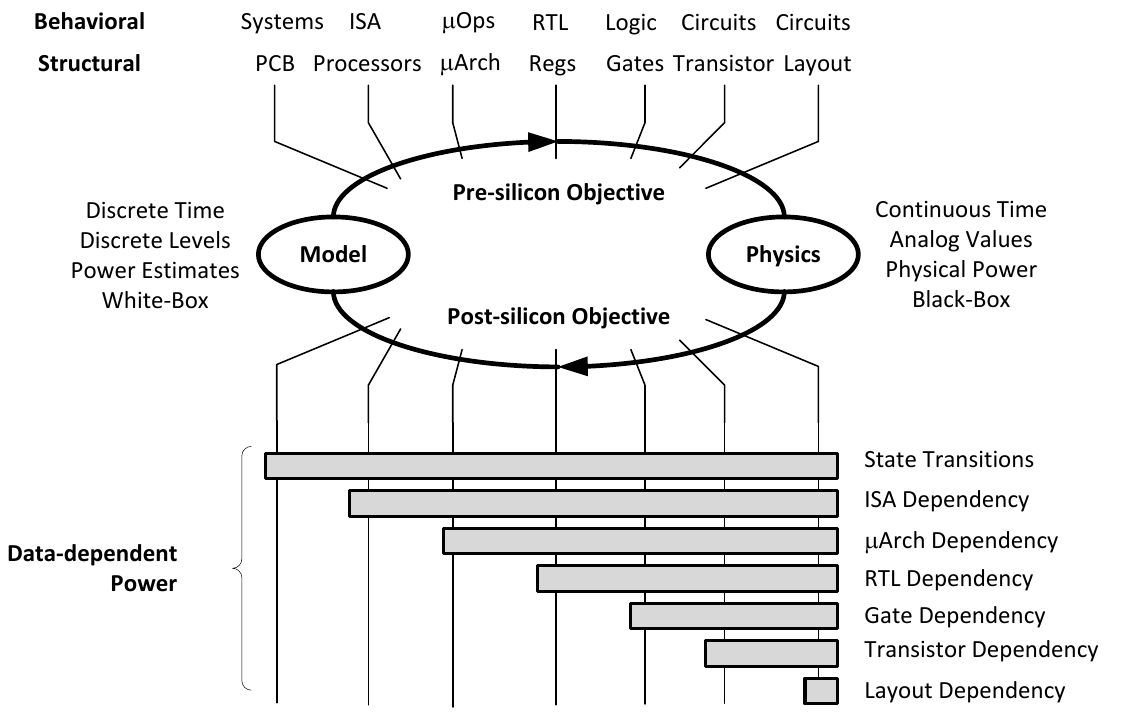}
\caption{Side-channel leakage modeling requires abstraction from the true physical source of side-channel leakage, thereby also abstracting some sources of side-channel leakage. Pre- and post-silicon side-channel leakage modeling both aim at building a model that reflects the true physical source of side-channel leakage, but approach the problem from opposite sides.\label{fig:synopsys}}
\end{figure}

Our main classification of tools in pre-silicon and post-silicon for leakage detection, summarized in \cref{fig:synopsys}, is based on the source of information for building the model which the tool uses.
There are two primary sources for such information at a high level: measurements taken from the device and the device's design specifications.
These two sources determine the two main axes along which we classify the tools.

The first axis specifies the relationship between the model and the hardware.
In all tools, the model aims to predict the leakage from the device. 
However, \emph{post-silicon} tools build the model, at least partially, based on measuring the device and using this measurement to predict future behaviour.
In contrast, \emph{pre-silicon} tools use information about the device's design to predict leakage without observing the actual device.
This distinction is not arbitrary.  Pre- and post-silicon tools serve different purposes and have different capabilities.
Pre-silicon tools aim at detecting leakage early, before the production of the device, allowing the designer the opportunity to modify the design before investing in manufacturing.
Post-silicon tools, in contrast, start from a given device and aim to determine the leakage that a change in condition, typically a new software, will cause.
Pre-silicon tools typically operate under a white-box scenario, where the model is created from the device's design.
They, therefore, include information typically only available to the manufacturer or trusted clients.
On the other hand, post-silicon tools operate under a black- or gray-box scenario, where the tool operator does not have the entire device description.
However, because these tools have access to actual measurements, they can detect leakage that is not apparent from the design documents.

The second axis we use is the level of abstraction of the model.
Development of a hardware device typically proceeds along a sequence of refinements, commonly captured in the Gajski-Kuhn Y-chart~\cite{GajskiK83}.
With each refinement, the abstraction level decreases, and more details about the target device are generated.
The level of abstraction used for building the model has significant implications on the tool's capabilities.
The more refined the model is, the more leakage it can detect~\cite{Buhan+20}.
This is particularly relevant for pre-silicon tools, where modeling at one level cannot detect leakage caused by features that are only determined at more advanced levels of abstraction.
For example, pre-silicon tools that model at the Register Transfer Level (RTL) cannot detect \emph{glitches}, which depend on timing information only available at the logic level abstraction.
Correspondingly, modeling at a high level of detail requires both access to the design documents and a considerable investment of computational and time resources.
\cref{fig:synopsys} shows the primary abstraction levels in the design of hardware devices and the types of data dependencies apparent at each level.
Finally, we note that because post-silicon tools also draw on information measured from the concrete device, such tools can detect leakage at a lower level of abstraction~\cite{ROSITA_2019,ASCOLD_2017}.

Through the classification and analysis of the published tools,  we identify the potential and the challenges we face when searching for appropriate solutions.
There are significant differences between existing tools, and some problems cross-cut across the domain. In particular, most tool proposals include an evaluation of the effectiveness of the tool.
However,  many of these evaluations are not transferable across tools. Thus, it is impossible to compare the effectiveness of tools within the domain.
The problem is exacerbated by the need to satisfy multiple aims, including simulation and detection accuracy, ease of use, and computational complexity.

We believe the work presented in this paper is of interest for two groups with distinct goals and challenges. The first group consists of hardware designers creating side-channel secure designs and side-channel hardened implementations at the pre-silicon stage. The second group are security researchers implementing a hardened side-channel cryptographic algorithm at the post-silicon stage.
Arguably,  it might be stated that such tooling could help an adversary interested in extracting a key from a specific device. However, we believe that the risk is somewhat limited.
The tools we describe aim at assisting designers in identifying the root cause of leakage. 
Attackers are less interested in the cause of a leak and are more focused on recovering the key. 
In summary, the contributions of this work are:

\begin{itemize}
  \item We investigate an emerging research area on automated tools for side-channel leakage detection and propose a system for classifying such tools. (\cref{sec:synopsys}).
  \item We survey and analyze post-silicon (\cref{sec:post_silicon}) and pre-silicon (\cref{sec:pre_silicon}) tools, identifying both achievements and challenges.
  \item We explore the cross-cutting questions by evaluating tools across the domain. (\cref{sec:Evaluation})
  \item We identify open problems and directions for future research on the design of automated tools for leakage detection. (\cref{sec:open_questions})
\end{itemize}

\begin{figure*}[t]
\centering
\includegraphics[width=0.9\textwidth]{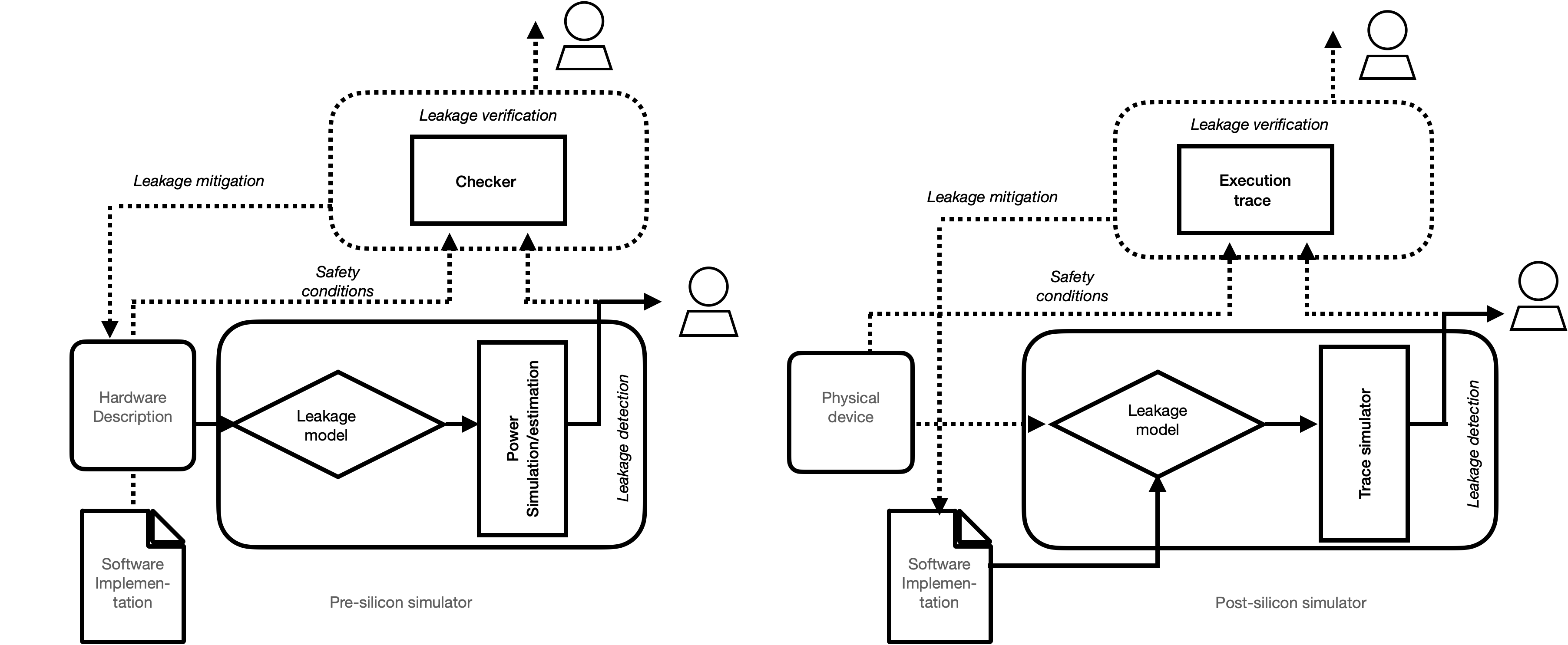}
\caption{ 
High-level architecture of a pre and post silicon simulator.   With continuous line, we denote the essential components required for leakage detection and, with a dotted line, the optional but helpful functionality of verification and mitigation.  One of the evident difference between the two is the object of the simulation. Post-silicon simulators, see \cref{sec:post_silicon},  are used for securing software implementations; when present, the physical target generates a more precise leakage model.  Pre-silicon simulators, see \cref{sec:pre_silicon}, are used to secure a hardware target, which in some cases will be tested in combination with a software implementation. }
\label{fig:anatomy}
\end{figure*}

\section{Background}
\label{sec:Background}
Throughout this paper, we use the term \emph{simulated trace} to denote a time series generated with a commercially available tool as part of an EDA toolchain. We use the term \textit{estimated trace} to refer to a time series derived through a custom-defined, so-called ``selection function'', which using a leakage model, maps sensitive values to specific predictions, typically used to estimate the power or the EM consumption in the context of side-channel evaluation. Finally, we use \textit{measured trace} to denote the traces acquired from a physical device with the help of an oscilloscope and EM probes. The term \textit{leakage simulator} stands for a device emulator connected to a leakage model, which creates a simulated or estimated trace. 

\textit{Leakage  detection} seeks  evidence  of  sensitive data dependencies in the measured traces. The tools typically used  for  leakage  detection  are  hypothesis  testing.  Due  to  hypothesis  testing’s  intrinsic  nature,  it is only possible to confirm the leaks’ presence (and not the absence).  \textit{Leakage  verification} aspires  to  identify  the  cause of  a  leak.  The  most  straightforward  way  to  verify  a  leak  is to  exploit  it  using  an  attack-based  evaluation.  Alternatively,  it  is   also  possible  to   specify  a   set  of  rules   that  violate the  algorithm’s  safe  run  assumptions,  e.g.,  register  reuse  by mask  shares  from  the  same  family.  To  determine  the  leak’s cause,  a  careful  investigation  of  the  hardware  and  software’s internal  working  is  required.  Finally,  \textit{leakage  mitigation} will remove the cause of the leak.

\subsection{Differential Power Analysis}

The threat of Differential Power Analysis (DPA) attacks became evident already in the '90s by demonstrating how
cryptographic keys can be extracted from embedded devices, e.g., smartcards, by merely observing a side-channel such as
timing or power~\cite{Koc96, KocherJJ99}.  A DPA attack that is using Pearson correlation for the key recovery is often called Correlation Power Analysis (CPA).
Commonly used metrics to evaluate the DPA attack's performance are \textit{guessing entropy} and \textit{success
rate}. The former represents the averaged key rank computed for all key candidates and the success rate of a side-channel
the attack is defined as the probability that the secret target key is ranked first among all key guesses by a score vector. 
\textit{Signal-to-Noise ratio} (SNR)~\cite{SNR_Mangard_2004} allows designers of cryptographic algorithms to verify that the combination of countermeasures they have chosen to implement in their device provides the required resistance against DPA attacks. 

\subsection{Profiled Attacks}

There exists a difference in the approach taken by DPA attacks to another class, so-called \textit{profiled attacks}. The latter, often
referred to as a two-stage attack, assumes an ``open'' device (or a copy of it) for the profiling stage in which most of the attack
work is done, while the critical recovery stage requires only a few
measurements or, in some cases, a single measurement~\cite{WeissbartCPB20}.
Examples include template attack~\cite{ChariRR02},
stochastic models~\cite{SLP05}, and recently machine learning-based
attacks~\cite{lerman2015template, maghrebi2016breaking}.
\subsection{Leakage Assessment methodology}
\textbf{Test Vector Leakage Assessment} (TVLA)~\cite{Goodwill_TVLA} is one of the most popular methods for leakage detection due to its simplicity and relative effectiveness. It is based on statistical hypothesis testing and comes in two flavors: \textit{specific} and \textit{non-specific}. The 'fixed-vs-random' is the most common non-specific test and compares a set of traces acquired with a fixed plaintext with another set of traces acquired with random plaintext.  In the case of a specific test, the traces are divided according to a known intermediate value tested for leakage. In both cases, Welch's two-sample $t$-test for equality of means is applied for all samples in the measured trace set.  A difference between two sets larger than a given threshold is taken as evidence for the presence of a leak. TVLA is creatively applied in many forms for both pre-silicon and post-silicon evaluation. 

\section{Pre- and Post-silicon Modeling of Side-channel Leakage}
\label{sec:synopsys}

A fundamental property of power-based (or other) side-channel leakage is that its origin is a byproduct of the physical implementation of computations. While the leaked information may relate to any higher level form of computing -- such as cryptographic software -- the observation of power-based side-channel leakage requires access to the physical implementation of the cryptographic computations. Nevertheless, there are ample opportunities for such observations, including remote observation of power consumption. Recent works such as PlunderVolt~\cite{DBLP:journals/ieeesp/MurdockOGBPG20} and Screaming Channels~\cite{DBLP:conf/ccs/CamuratiPMHF18} have demonstrated that this remote access can be implemented in a multitude of ways.

The physical effects of computing  are not harmful to computer system security by themselves; such effects only become side-channel leakage when an association can be made between the physical side-channel leakage and a high-level property in the computation stack.  Hence, fundamental to every side-channel attack is the association of a high-level model with its physical implementation.  The accuracy of this association determines the success of the attack. Therefore, side-channel leakage modeling is of great interest because it builds insight into the link between physical side-channel leakage and a high-level property.

The complexity of side-channel leakage modeling stems from the fact that computing infrastructures are implemented over many layers of design abstraction. The physical effects of computing are hidden by choice,  for design efficiency and convenience.  At higher levels of abstraction, it is difficult to understand or anticipate the physical effects of side-channel leakage. There are two distinct flavors of the side-channel leakage modeling problem, and we identify them as pre-silicon modeling and post-silicon modeling. Pre-silicon modeling predicts physical side-channel leakage based on high-level design information such as detailed hardware descriptions. Pre-silicon modeling is a task faced by the hardware designer of a secure chip. Post-silicon modeling estimates the properties of the higher abstraction levels in the computation stack based on the physical observation of side-channel leakage. Post-silicon modeling is the task faced by the software programmer who aims at writing side-channel secure software for an off-the-shelf processor.

\cref{fig:synopsys} is a synopsis that summarizes the ideas of the following paragraphs. Pre-silicon modeling typically operates under a white-box assumption, where the implementation details of the computing stack are known, even when the exact path towards physical implementation has not been completed. The Gajski-Kuhn Y-chart defines the typical abstraction levels in computer hardware and identifies a computing system's behaviour and structure. At the lowest,  most detailed abstraction level, the structure is captured by the chip layout, which is then abstracted into less detailed structures such as a transistor netlist,  gate netlist,  register-transfer description,  processor block diagram, up to system-level structural descriptions. Similarly, analogue continuous-time circuit semantics is captured at the most detailed level, which is then abstracted into logic levels and clock cycles, register transfers, processor instructions, up to system-level task descriptions. The computer chip under construction is decomposed into each of these behavioural and structural forms until a design is obtained at the most detailed level, ready for tape out. With pre-silicon side-channel leakage modeling, a designer tries to predict the physical computer chip's side-channel leakage using any design descriptions available.

Post-silicon modeling operates under a black-box assumption, meaning that the implementation details of the computing stack are only partially known, even as a physical artifact (such as a processor) is  available. Post-silicon modeling establishes a high-level model that enables reasoning about side-channel leakage; this high-level model has to be expressed at the abstraction level required for side-channel analysis. Post-silicon modeling is helpful when designing programmable and configurable systems. Post-silicon modeling works along the same behavioral and structural abstractions as pre-silicon modeling.

The bottom of \cref{fig:synopsys} identifies some of the most important sources
of data-dependent power dissipation. In the broadest sense, any data-dependent power dissipation can result in power-based side-channel leakage. An important observation in the abstraction levels of computation is that each level adds its own form of data-dependent power dissipation to the overall data-dependent power dissipation. The physical measurement of side-channel leakage is the sum of all these effects in addition to measurement noise. Models at higher design abstraction levels therefore lose some details of the side-channel leakage,  and hence every design abstraction level deserves verification of side-channel leakage properties. Some examples of known sources of power-based side-channel leakage given below serve to illustrate this point.

\begin{itemize}
\item State transitions, in the form of intermediate computation results, are the most traditional source of side-channel leakage, and they are often directly used in formulating a side-channel attack. 
\item When software executes on a processor, data-dependent power dissipation can occur because of dependencies in the instruction set - such as the dependency of power consumed by an instruction on similar operands. 
\item Similarly, the micro-architecture can add additional register transfer level dependencies when operands travel in the internal processor architecture. 
\item At gate level and below, detailed implementation effects lead to glitches, a non-linear source of data-dependent power dissipation.
\item Transistor circuits dissolve to convenient logic threshold levels from 0 and 1 into an analog range of voltages. Circuit effects such as static power leakage, which is invisible at the gate level, leads to data-dependent power dissipation.
\item Finally, the physical geometries of the layout lead to interactions between neighboring wires and parasitic effects in the power distribution, both of which can cause data-dependent power dissipation.
\end{itemize}

To be useful, a leakage model must be ``correct'' to accurately reflect reality and be informative to be useful for key recovery.  As in~\cite{ASCOLD_2017}, we distinguish between \textit{value} and \textit{distance}-based leakage models. The leakage model is value-based if it takes as arguments the set of intermediate values of a cryptographic algorithm. Typical examples include the popular Hamming-weight (HW), Hamming-distance (HD), identity model (ID), or least-significant bit (LSB). A leakage model is \textit{distance-based} if it takes as parameters any pairwise combination of the intermediate values~\cite{ASCOLD_2017}. Barthe et al.~\cite{Barthe_2021_Masking_in_fine_grained_leakage_models} give three examples for distance-based leakage, as follows: (1) \emph{transition} leakage effect, such as a generalized Hamming distance leakage model; (2) the \emph{revenant} leakage effect, where sensitive data from past executions may come back and influence the current instruction; and (3) the \emph{neighbouring} leakage effect, which captures the event where accessing or processing of a data storage unit, may trigger a leak from a seemingly unrelated data storage unit~\cite{Barthe_2021_Masking_in_fine_grained_leakage_models}.  In the rest of the paper, we will use the following categories for leakage models: 
\begin{description}[font=$\bullet$~\normalfont\scshape\color{red!50!black}]
\item [Black-box] typically value-based, there is no information about the placement of the circuit elements or the routing signal between them is determined using little or no specific information about the physical target on which the algorithm will run. This typically corresponds to ISA level information or architecture details - in the form of high-level code assembly instructions or low-level machine code. This leakage model is sometimes referred to as \textit{behavioral level simulation}. 
\item [Gray-box] combines the value with partial distance-based leakage models. Partial micro-architectural details of the physical target are used to derive the leakage model, obtained by either reverse engineering the micro-architecture or profiling the physical target using specialized equipment. 
\item [white-box] Full knowledge of the target, access to RTL gates, or layout description of the target. We include in this model both \textit{gate-level} simulations, where the instantaneous power consumption of a circuit is modeled as "toggle count", calculated as the (weighted) sum of the number of transitions in each gate, and the \textit{transistor level} simulations where the power consumption is modeled as a set of differential equations. 
\end{description}

\section{Post-silicon tooling: State of the Art}
\label{sec:post_silicon}
 
The most basic functionality of a post-silicon simulator is that of leakage detection, Figure~\ref{fig:anatomy} (left). 
There are two main approaches for this task. The first, and by far the most common,   mimics leakage detection in real traces, by first generating a set of simulated traces and then applying a leakage detection method (e.g. TVLA~\cite{Goodwill_TVLA}).  Generating realistic simulated traces depends to a great extent on the amount of information available about the target end device. The second approach  uses \textit{safety checks}  to identify undesired interactions of sensitive variables. To verify the leakage of an implementation and to identify the location of the leakage, the structure or architecture of the target device must be known~\cite{ASCOLD_2017}.  Leakage mitigation requires the ability to modify the target device by reprogramming or reconfiguring~\cite{ROSITA_2019}.

\cref{tab:post_silicon_simulators} presents a taxonomy of tools available for post-silicon side-channel   evaluation,  listed in chronological order.  Based on their capabilities,  we classify the tools in three categories:  detection~(D),  verification~(V) and mitigation~(M), which we discuss subsequently in detail.  The supported leakage model (LM) is mentioned explicitly as it has an important impact on the effectiveness of the tool. At one end of the spectrum, the common black-box leakage models, such as Hamming weight or Hamming distance,  can be applied independently of the intended physical target. They only require a high-level description of the implementation and give a rough estimate of the actual power or EM consumption.  This is enough to model the data dependencies during the execution and may give valuable insights into value-based leakage. At the other end of the spectrum,  gray-box leakage models learn from the intended target's behavior by acquiring traces from the actual implementation, making the analysis specific to a particular sequence of instructions.  

The amount of information and the degree of control of the end-device available when building the simulator (end-device), determines the capability of the tool (D,V, or M) and has an impact on how fined-grained is the leakage model of the end-device. On the down-side,  the more information a simulator captures about the target, the less portable to other architectures it will be.  

Some tools are designed without a physical end-target~\cite{Inspector_SCA,Reparaz_2016}. As such, they are not necessarily useful for post-silicon evaluations \textit{only}.  We chose to list them in this category as these tools can be used for early design stages of software implementation, as the typical use case for  post-silicon evaluation tooling. As masking is one of the key countermeasures for software implementations, we find it important to specify whether a tool has been demonstrated on  a masked implementation (Masking).   

\begin{table*}[ht]
\caption{Tools for post-silicon side-channel evaluation (chronological order) }
\centering
\begin{tabular}{llclcccclc}
\toprule
Name & Year &LM & End-device & D & V & M &Masking & Side-channel& Open-Source \\
\midrule
PINPAS~\cite{PINPAS_2003}           & 2003 & \fullcirc      & smartcards    & \cmark & \nmark & \nmark & \xmark & power & \xmark \\

Inspector SCA~\cite{Inspector_SCA}     & 2007 & \fullcirc      & not relevant     & \cmark & \nmark & \nmark & \xmark & power & \cmark\$\\
Oscar~\cite{OSCAR_2009}           & 2009 & \fullcirc      & AT90XX, ATmegaXX    & \cmark & \nmark & \nmark & \xmark & power & \xmark \\

Debande~\cite{Debande_2012}         & 2012 & \halfcirc      & not specified  & \cmark & \nmark & \nmark & \xmark & power & \xmark \\
Gagnerot~\cite{Gagnerot_2013}           & 2013 & \fullcirc      & Risc-V( not specified)   & \cmark & \nmark & \nmark & \xmark & power & \xmark \\
SILK~\cite{SILK_2014}               & 2014 & \fullcirc      & ATmega328P    & \cmark & \nmark & \nmark & \xmark & power & \cmark \\
SLEAK~\cite{Sleak_2014}   & 2014 & \fullcirc      & \textit{ARM Cortex A8}  & \cmark & \cmark & \nmark & \cmark & register access & \xmark \\
Reparaz~\cite{Reparaz_2016}         & 2016 & \fullcirc      & not relevant          & \cmark & \nmark & \nmark & \cmark & power & \xmark \\
SAVRASCA~\cite{SAVRASCA_2017}       & 2017 & \fullcirc      & ATMega163     & \cmark & \cmark & \nmark & \cmark & power & \cmark \\
ASCOLD~\cite{ASCOLD_2017}           & 2017 & \halfcirc      & ATMega163     & \nmark & \cmark & \cmark & \cmark & ILA   & \cmark \\
ELMO~\cite{ELMO_2017}               & 2017 & \halfcirc      & ARM Cortex M0 & \cmark & \nmark & \nmark & \xmark & power & \cmark \\
$\text{ELMO}^*$~\cite{ROSITA_2019}  & 2019 & \halfcirc      & ARM Cortex M0 & \cmark & \nmark & \nmark & \cmark & power & \cmark \\
ROSITA~\cite{ROSITA_2019}           & 2019 & $\text{ELMO}^*$& ARM Cortex M0 & \cmark & \cmark & \cmark & \cmark & power & \cmark \\
EMSIM~\cite{EMSIM_2020}& 2020 & \halfcirc & Risc-V(custom)&  \cmark & \nmark & \nmark & \nmark & EM& \xmark \\
\bottomrule \\

\multicolumn{10}{c}{We use \fullcirc{} to represent a black-box leakage model (LM); \halfcirc{} to represent a gray-box model. We tick the box for masking} \\
\multicolumn{10}{c}{
for the tools that report a case study involving a masked algorithm.}\\
\end{tabular}
\label{tab:post_silicon_simulators}
\end{table*}

\subsection{Leakage detection at post-silicon stage}
\label{sec:detection_post_silicon}

In the following subsections, we describe post-silicon tooling for leakage detection, leakage verification, and leakage mitigation. For each tool, we highlight achievements and challenges.

\noindent
\textbf{\textit{Achievement: Generic framework for modeling of micro-architectural details in a black-box model}. }
Debande et al.~\cite{Debande_2012}, the first to point out the significance of deriving realistic leakage models, propose the first gray-box trace simulator. The simulator uses stochastic modeling to fit a function of state bits and state transitions. It starts from a fixed model and estimates the state transitions for each bit in the target register.  

We consider ELMO~\cite{ELMO_2017} to be the first truly gray-box simulator for the ARM-Cortex M0/M4 family. It brings two remarkable innovations. The first one is a portable framework for building a leakage model rather than estimating the coefficients for a fixed model, as is the case for stochastic modeling.  ELMO achieves this by considering the contribution of a parameter before deciding to include it in the model. The second is the extension of the model to support \textit{sequence dependency}. The key observation is that different instructions' power consumption depend on the instructions executed before that~\cite{Tiwari_Instruction_level_power_analysis}. ELMO is instruction-accurate, which has the advantage of easily allowing the identification of a leaky instruction. When modeling the power consumption, the authors disregard the high registers (\texttt{r8--r15}) because those are only used for fast temporary storage. Following a cluster analysis to group ``similar'' instructions (i.e., which leak information in the same way), the authors identify five groups, which interestingly also correspond to the same processor component: ALU instructions in one group, shift instructions as another group, load and stores that interact with data as two or more groups, and multiply instruction with a distinct profile due to its single cycle implementation.  The authors find a remarkable consistency in the data-dependent leakage of different physical boards. The only downside for extending the proposed framework to other architectures is the amount of human effort which has to be put into it. ELMO is open-sourced and publicly available. 

ELMO*~\cite{ROSITA_2019} improves the leakage model of ELMO by capturing interactions that span multiple cycles.  ELMO~\cite{ELMO_2017} is augmented to account for the storage elements, which play a critical role for the security of masked implementations. A novel feature of ELMO*~\cite{ROSITA_2019} is a battery of small code sequences which can be used to systematic  highlight the interaction of instructions via storage elements. The idea for finding the hidden storage elements is generic and can be applied for any other architecture. 

\noindent
\textbf{\textit{Achievement: EM simulation at post-silicon stage}.} While both EM and power are important for SCA evaluations,  modern micro-controllers,  with multiple power domains,  can be immune to power side -channels but can leak in the EM domain.  
EMSIM~\cite{EMSIM_2020} is the first EM simulator built for a custom 32-bit base Risc-V implementation. 
The simulator supports data and instruction dependent activities and micro-architecture effects such as pipeline stalls, cache miss, and misprediction. 
A comparison between the simulated and measured EM signal is performed to determine the simulated signal's quality.  The result shows the two signals to be very close. To reduce the number of instructions that need to be fitted, the authors do perform clustering of the power consumption of the instructions and observe that the Risc-V ISA can be clustered into seven categories when the instructions have similar operands. 

Interestingly, the same phenomena was  identified and used by ELMO~\cite{ELMO_2017} to simplify the modeling of the target. The manufacturing variability (same manufacturer, different physical boards) on the model accuracy was investigated. Although the authors detected a slight shift in the clock frequency for different boards, the conclusion is that this shift has no impact on the accuracy of the simulator. Furthermore, the model accuracy was explored as a function of board variability (same core, different manufacturer). The conclusion was that for different designs, only the baseline amplitude and activity factors should be retrained.  

\noindent
\textbf{\textit{Challenge: EM modeling requires access to design details.}}  The authors of EMSIM~\cite{EMSIM_2020} 
show that having access to micro-architectural details is critical for achieving good accuracy for EM simulations.

\noindent
\textbf{\textit{Challenge: Access to tools.}}  Although in recent years the standard practise is to open-source tools (as the authors are mostly from academia), for many of the earlier tools~\cite{PINPAS_2003,OSCAR_2009,Gagnerot_2013} we only have a description from the authors about the tools capability and innovations,  which in many cases provides limited information.

\noindent
\textbf{\textit{Challenge: The existing simulators target relatively simple architectures}.} As can be seen from \cref{tab:post_silicon_simulators} the most commonly targeted end-devices are ATMegaXX or ARM-Cortex M0,  which are simple, in-order,  single-core CPUs.  In addition, the simulators which target the micro-architecture of the design require significant effort to port to other architectures.  With no access to design information, the task of the designer is to reverse-engineer the micro-architecture details. The most notable exception is the tool called SLEAK~\cite{Sleak_2014} which is showcased on the ARMCortex A8,  a modern and complex processor.  To access the values of intermediate state,  SLEAK uses Gem5 as an open source- full system simulator.  Power consumption is modelled with a black-box leakage model.  

\noindent
\textbf{\textit{Challenge: Lack of integration with formal verification tools.}} 
Oscar~\cite{OSCAR_2009} is a power simulator tool,  designed for 8-bit Atmel AVR micro controllers,  constructed in a pure functional style (\texttt{OCaml} language) with the intention of integrating the power simulator tool into formal proofs of resistance.  The default supported leakage models are the black-box leakage models, but the tool allows the monitoring of successive microprocessor states or partial states (e.g., a register or memory).  Oscar is the only leakage detection tool in the post-silicon category which aims to bridge the gap between physical security and formal verification.

\subsection{Leakage verification at post-silicon stage}
\label{sec:verification_post_silicon}

\noindent
\textbf{\textit{Achievement: Verification at high  abstractions levels is effective}.}
After detecting the presence of a leak in an implementation, mitigating the leak requires discovering the cause of the offending instruction.  The tools which are capable of locating the leak must possess the capability of mapping a time sample in the power trace to the precise instruction responsible corresponding to that time sample.  The alternative to the tools which identify the cause of the leak is either an educated guess or trial and error.

The tool proposed by Reparaz~\cite{Reparaz_2016} can detect leakage in masked implementations of high-level code. The tool has a trace generation feature that uses a black-box leakage model.  For each time sample, the value of the processed variable is also recorded.  A fixed-vs-fixed test is used for leakage detection. As the tool records which variables correspond to the leaky sample, it is possible to locate the source of the leakage.  

SAVRASCA~\cite{SAVRASCA_2017} uses the tracing feature of the SimulAVR tool and is suitable for the analysis of code for the AtmelAVR family. The simulator can produce both power and execution traces.  To create power traces, the leakage model is computed during each memory unit access (available via the tracing feature of SimulAVR), making a difference between a write and read access. The separation allows for different leakage functions depending on the type of access (Hamming weight for reading and Hamming distance for writing). The simulator produces one power sample per executed instruction and does not consider the memory unit's address. 

\noindent
\textbf{\textit{Challenge: Mapping a time sample in a measured trace to the corresponding executed instruction is difficult}.}
The tools which can map a time sample in a measured trace to the corresponding instruction of the executed  code are limited and typically fall in two categories: 
either a machine emulator such as Qemu~\cite{QEMU}, Gem5~\cite{GEM5} or SimulAVR~\cite{SimulAVR} or specialized hardware,  such as JTrace Pro\footnote{\url{https://www.segger.com/products/debug-probes/j-trace/models/j-trace/}} for hardware which provides advanced debug probe supports the tracing features of ARM Cortex Cores.  Therefore,  verification depends on whether a machine emulator supports the board,  or the board has a tracing pin availa

\subsection{Leakage mitigation at post-silicon stage}
\label{sec:mitigation_post_silicon}

While verification tools still rely on a human expert to remove leakage, mitigation tools aim to apply the fixes automatically. 

\noindent
\textbf{\textit{Achievement: Generic code-rewrite for trace simulators.}}
ROSITA~\cite{ROSITA_2019} is a rule-driven code rewrite engine that patches the code automatically once leakage is detected.  ROSITA starts with a (masked) implementation of a cryptographic algorithm, cross-compiled to produce both the assembly and the binary executable.  A very compelling feature of ROSITA is that it extends an existing leakage detection tool,  ELMO~\cite{ELMO_2017}  to report instructions that leak secret information.  The new detection framework ($\text{ELMO}^*$), uses the binary file to detect leakage and identify the offending machine instruction,  ROSITA then applies a set of rules that replace the leaky instruction with an equivalent one (functionally) that does not leak. The process is repeated until no more leakage is detected. The ROSITA concept can be extended to other architectures. 

\noindent
\textbf{\textit{Challenge: Distance-based leakage is platform specific.}} ASCOLD~\cite{ASCOLD_2017} can verify code and mitigate leakage  by checking violations of the independent leakage assumption (ILA), responsible for reducing the actual security order of an implementation. The tool takes as input the assembly file of a masked implementation and a configuration file which describes the initial state of the system, e.g. the registers or addresses in the memory which contain the secret shares or sensitive variables. The output of the simulator is the line number and the rule that was violated by the program. The algorithm verifies for \textit{overwrite effects}, so that shares from the same family are not written in the same register, it checks the \textit{memory remenant effect} of the load/store instructions and \textit{neighbouring leakage} where an access of a register will cause a leakage in a unit elsewhere.  Unfortunately, these effects depend on the architecture and are only observable throughout an implementation.  As in the case of ROSITA~\cite{ROSITA_2019},  ASCOLD assumes a detailed description of the micro-architectural effects of the target board, which requires intensive effort to determine.

\section{Pre-silicon tooling}
\label{sec:pre_silicon}
 
The world of pre-silicon side-channel leakage verification tooling, while at first glance relatively rich (see \cref{tab:pre-silicon-tooling}), is limited by the fact that these tools are not public and the results cannot be reproduced. Additionally, the reported results consider a prototype chip design for reporting, which might significantly adapt to a complex chip design. While any measurable data-dependent power dissipation may be a source of side-channel leakage, there is a trade-off between the precision and the simulation speed. Higher abstraction levels (ISA, RTL) will offer quick power estimates. Still, they will miss SCA leakage sources, while lower abstraction levels (gate layout) consume more simulation time but are more precise. In the following, we summarize the simplifications made by the different tools to estimate the power or the EM consumption and the target used for modeling. \cref{fig:anatomy}~(right), a conceptualized architecture of a pre-silicon tool, will guide our narrative. 
 
\begin{table*}[th]
\caption{Tools for pre-silicon side-channel verification (chronological order)}
\centering
\begin{tabular}{llllcccccllc}
\toprule
Name&Year & Input &End-device (description)&LM&D& V& M&Masking&Target& Side-Channel&Open-Source\\
\midrule
NCSIM~\cite{NCSIM_2007}     & 2007 & gate   & SCARD~\cite{SCARD}                                  & S & \cmark & --     & --     & \xmark & CC    & power & \xmark \\
AMASIVE~\cite{AMASIVE_2013} & 2013 & RTL    & --                                                  & E & \cmark & --     & --     & \xmark & CA    & power & \xmark \\
MAPS~\cite{MAPS_2018}       & 2018 & ISA    & ARM CortexM3                                        & E & \cmark & \cmark & --     & \cmark & CA    & power & \cmark \\
KARNA~\cite{KARNA_2019}     & 2019 & layout & AES~\cite{Karna_TinyAES}, SIMON~\cite{Karna_Simon}  & S & \cmark & \cmark & \cmark & \xmark & CC    & power & \xmark \\
RTL-PSC~\cite{RTL_PSC_2019} & 2019 & RTL    & AES-GF~\cite{RTL_PSC_16}, AES-LUT~\cite{RTL_PSC_17} & S & \cmark & --     & --     & \xmark & CC    & power & \xmark \\
PARAM~\cite{PARAM_2020}     & 2020 & gate   & Risc-V(ShaktiC)                                     & E & \cmark & \cmark & --     & \xmark & ED    & power & \xmark \\
COCO~\cite{COCO_2020}       & 2020 & gate   & Risc-V                                              & --& \cmark & \cmark & --     & \cmark & CA+ED & power & \cmark \\
ACA~\cite{ACA_2020}         & 2020 & gate   &Risc-V(LEON3)                                        & S & \cmark & --     & --     & \xmark & CC    & power & \xmark \\
SCRIPT~\cite{SCRIPT_2020}   & 2020 & gate   & AES-GF~\cite{RTL_PSC_16}, AES-LUT~\cite{RTL_PSC_17} & -- & \cmark & \cmark & --     & \xmark & CC    & power & \xmark \\
CASCADE~\cite{CASCADE_2020} & 2020 & gate   & ASIC (custom)                                       & E & \cmark & --     & --     & \xmark & CC    & power & \cmark \\
\bottomrule\\
\multicolumn{12}{l}{\textit{Input} specifies the abstraction level for the input to the simulator, for the end-device specified by the column \textit{End-device}.  For the leakage}\\
\multicolumn{12}{l}{model\textit{(LM)} we have two options: simulated power (S) or estimated power(E.). Columns (D,V,M and \textit{Masking}) are defined as in \cref{tab:post_silicon_simulators}}\\
\multicolumn{12}{l}{The \textit{Target} column describes what is being simulated: the crypto core (CC), the cryptograpfic algorithm(CA) or the end-device(ED).}\\

\end{tabular}
\label{tab:pre-silicon-tooling}
\end{table*}

\subsection{Leakage detection at pre-silicon stage}
\label{sec:detection_pre_silicon}

\noindent
\textbf{\textit{Achievement: Power estimates at different abstraction levels speed-up the generation of power signals.}} 
NCSIM~\cite{NCSIM_2007} is the first white-box simulator to estimate DPA resistance at the gate level. It neglects the static power consumption, but it can model glitches and early propagation when timing information is added to the model. The simulator supports several power estimation functions. The simplest one is \textit{transition counting} (each time the signal changes its logical state, the power consumption, at the current point, is increased by one). This information is present in the VCD files, and a Matlab toolbox was used to estimate the power trace.  A more refined power model is the \textit{random transition weighting}, which captures the fact that the load capacitances are not identical for every gate (the experiments are performed on a dual-rail precharged logic style) implemented by adding a random weighting to each transition. Finally,  \textit{back-annotation of the transition weighting} can also be added by extracting the full-chip layout's parasitic information. 
In terms of speed,  NCSIM reports that a transistor level simulation of an internal MOV operation including the initialization phase of the core, has taken about 10 hours vs, the logic simulation that finishes in a few minutes.  
	
PLAN/PARAM~\cite{PARAM_2020} estimates the power consumed by a module as an aggregation of the power consumed by all signals present in the module.  The assumption to support this choice is that the power consumption of a  $k$-bit signal is proportional to its Hamming weight. The benefit of this approach is that the whole Shakti-C processor's evaluation takes about 5 hours compared to a post-and-place route simulation that would take a complete month.

RTL-PSC~\cite{RTL_PSC_2019} estimates the power profile of a hardware design using functional simulation at the RTL level. To ensure a fast framework,  the power profile is estimated based on the number of transitions using the Synopsys VCS tool.  Compared to state-of-art, RTL-PSC claims two advantages. The first is the ability to quantitatively and accurately assess power side-channel leakage  and the second is speed. The evaluation time for AES-GF is 43.6 minutes and for AES-LUT for about 24 minutes. The same evaluation at the gate level would take about 31 hours, while the authors estimate it would take more than 1 month at the layout level. 

ACA~\cite{ACA_2020} uses a gate-level model for a target design, which is typically available after logic synthesis, as well as a side-channel leakage model. The latter leakage model is common in DPA attacks. The objective of ACA is to identify the gates in the design that are contributing the most to the selected side-channel leakage model. ACA introduces the Leakage Impact Factor (LIF), a numerical score that reflects the relative contribution of a single gate to side-channel leakage. The authors demonstrate for several different application scenarios (an AES engine, a Sparc-V CPU) that only a handful of gates can be identified as majority contributors to side-channel leakage. This leads to the mitigation strategy of selective replacement, in which only those gates with high LIF are substituted and protected by side-channel resistant versions.

CASCADE~\cite{CASCADE_2020}, a white-box simulator that aims to speed up the time to market and reliability of the secure design uses an  extended version of the Hamming Distance model,  named the \textit{Marching-Stick Model (MSM)} to model power consumption.  MSM is a generic model that captures the asymmetry between rising and falling edges, unlike the simple toggle counting.  When the tool is used to check second-order security using both the marching stick leakage model and the PrimeTime with PX, we see that both results indicate the presence of second-order leakage, as expected. The second use case is the Boyar-Peralta AES S-box~\cite{Cascade_13,  AES_DaemenR20}which was found to be leaky~\cite{Cascade_39}. To demonstrate the presence of the mentioned vulnerability, the setup used 10 million traces, but it takes CASCADE only 30 min to find the indicated vulnerability in the gate-level netlist. The third analysed S-box implementation is an in-house implementation of a PRESENT S-layer in WDDL,~\cite{DBLP:conf/ches/TiriHHLYSV05}  a dual-rail logic style. The analysis is done at both gate-level netlist and place and route, using both the simulated power model with PrimeTime and estimated power with the marching stick model. In all cases, no leakage is detected.  

\subsection{Leakage verification at pre-silicon stage}
\label{sec:verification_pre_silicon}

\noindent
\textbf{\textit{Achievement: Formal verification at gate level.}}  Both  SCRIPT~\cite{SCRIPT_2020}  and COCO~\cite{COCO_2020} allow formal verification at gate level.  However while SCRIPT targets crypto cores,  COCO is aimed at formally verfying a masked software implementation on a given hardware platform. The approach used for the formal proof, described below, is also different.  SCRIPT~\cite{SCRIPT_2020} takes as input a gate-level description of a crypto core,  and a \textit{target function},  which can be a potential target for side-channel attacks if it satisfies the following four properties:
\begin{itemize}
\item a function of the secret,
\item a function of controllable inputs,
\item a function with confusion property,
\item and functions with the divide-and-conquer property. 
\end{itemize}
The target registers (which store the target functions' output values) are identified using information flow tracking. To estimate the power consumption, SCRIPT uses a vectorless power estimation technique, which requires the verification engineer to define the signal probability (the percentage of the analysis when the input is driven at high logic levels) and the toggle rate (the rate at which the net or logic element switches compared to its input) of the primary input ports. The vectorless power analysis can be performed using PrimeTool (Synopsys) or XPE (Xilinx), which returns the total estimated power for the design.  COCO~\cite{COCO_2020} allows security proofs at gate level for the execution of masked implementations. The proofs are done in the time-constrained probing model (proposed in the same paper), which simulates the hardware of pipelined circuits. As a first step, the masked assembly implementation is executed on a given CPU hardware design, the result being a \textit{trace execution} which contains the concrete values for all CPU control signals in each clock cycle. The location of the registers and memory cells which contain shares of sensitive values are annotated.  Next, COCO uses correlation  sets and a SAT solver to find the exact gate and execution cycle where the implementation leaks.   In essence,  COCO verifies adherence to the following two design principles: first, that shares of the same secret must not be accessed within two successive instructions and second, that a register or memory location which contains one share must not be overwritten by its counterpart. 

\noindent
\textbf{\textit{Achievement: Open-source tooling.}}  
The first open-source verification tool  is 
MAPS~\cite{MAPS_2018}, a power simulator for the ARM Cortex M3 series. It takes in assembly code and targets pipeline leakage, as they combine operand values from consecutive instructions. For identifying power leakage, it uses the fixed vs-random t-test~\cite{Goodwill_TVLA}.  Using information from an ARM Cortex M3 HDL file,  the cause of a leak, the registers related to the data path are isolated and traced. To simplify the tracing and reduce the resulting power trace, the operation is restricted to registers which deal with sensitive values; as such, the \texttt{r15} which holds the program counter is ignored. Furthermore, the three ALU registers are discarded as they are used for multicycle instructions, which are not commonly used during crypto algorithms.  MAPS is not cycle-accurate and traces only registers of the ARM Cortex M3 core, other registers located outside the core are not considered.  While both MAPS and  COCO~\cite{COCO_2020}, the two open-source verification tools,  can be used to verify masked software implementations,  MAPS supports only the ARM Cortex M3 platform, while COCO can handle  any given netlist. 

\noindent
\textbf{\textit{Achievement: SCA resilience for non-cryptographic designs.}}  
PARAM~\cite{PARAM_2020} is a microprocessor design hardened for side-channel resistance. It is a trace simulator, but it features a Power attack Leakage Analyzer (PLAN) module, which works on the RTL source code to identify the target microprocessor's leaking module. The running example is the open source of the Shakti-C Risc-V processor. The processor is represented as a netlist of functional modules such as the main pipeline, the ALU unit, data cache, instruction cache, etc. For a given module, leakage is estimated from the signals (wires and registers) associated with the module.  Once the power consumption is estimated, SVF (Side-Channel Vulnerability Factor)~\cite{SVF_2012},  is used to calculate the leakage. The authors do mention among the caveats that PLAN can only capture linear leakage and leakage due to dynamic power consumption (also the most exploited one in side-channel attacks). 

\noindent
\textbf{\textit{Challenge: Differentiating the merits of the tools is difficult.}} If we compare the architectures of the tools, \cref{fig:anatomy}~(right),  we notice that SCRIPT and COCO use the user input to define safety conditions for the underlying architecture. To determine the presence of a leak, MAPS and PLAN/PARAM employ empirical leakage detection strategies,  t-tests~\cite{Goodwill_TVLA} and SVF~\cite{SVF_2012} respectively.   Furthermore, if we compare the input of the simulator, we observe that while MAPS and COCO target, a masked software implementation, SCRIPT aims at the verification of crypto cores and PLAN/PARAM aims to secure the end-device or non-cryptographic implementation.  If we explore the dimension of security guarantees,  SCRIPT and COCO aim for a formal proof, while MAPS and PLAN/PARAM take an empirical testing approach.

\subsection{Leakage mitigation at pre-silicon stage}
\label{sec:mitigation_pre_silicon}

\noindent
\textbf{\textit{Achievement: Leakage mitigation tool at layout level.}} KARNA~\cite{KARNA_2019} identifies vulnerable gates in the design and then re-configures them.  The chip is partitioned into small cells, and a TVLA assessment is done for each cell. To estimate the power consumption of a gate, KARNA uses commercial tooling, and the tool reveals leakage specific to a given area. 

\noindent
\textbf{\textit{Challenge: Removal of leaks are done empirically.}} KARNA~\cite{KARNA_2019} identifies side-channel security from the netlist by computing TVLA scores at the gate level. The assumption is that not all gates on the netlist contribute equally to the side-channel leakage. To remove a vulnerable region, KARNA first determines if the gate is critical (if it can not undergo any more configurations). If not, it will replace the vulnerable gate with its next low-power configuration.  KARNA will also optimize the gate parameters such that the overall security of the design improves while keeping the design requirements such as area, power, and delay.

\section {Evaluation criteria}
\label{sec:Evaluation}
All research contributions to put forward a tool for leakage detection, evaluation, or mitigation typically contain a part dedicated to the tool's validation and experimental results.  This is important as it demonstrates the practical value of the tool in identifying and removing side-channel leaks.  In this section, we examine the different metrics used to determine how the leakage simulator's output can be used for developing a side-channel hardened target.  Among the presented evaluation techniques, we identify four distinct groups:
\begin{enumerate}
\item  Comparison between  simulated/estimated and reference traces,  relevant mostly for trace simulators.  
\item  Evaluations of \textit{leakage model's quality,}  most often through comparing it to a simpler model. 
\item Evaluation by case studies,  where the simulator is used to find  and/or fix side-channel leaks. 
\item \textit{Usability measures} explore related benefits for using a simulator, typically by comparing the performance of the tool with either a measurement setup for the post-silicon simulators or the power simulation techniques for the pre-silicon simulators. 
\end{enumerate}   
It is important to mention that most simulators are evaluated using a subset of the groups mentioned above.

\subsection{Metrics for evaluating the output of leakage simulators} 

The question answered by the metrics we place in this group can be summarized as: \textit{how close is the output of a leakage simulator to the reference traces?}  The implicit assumption is that the closer the leakage simulator's output to the reference traces, the more it can be trusted. Different boards exhibit different behaviours~\cite{Unai_When_similarities_are_taken_for_granted, Bhasin_Mind_the_portability} that may cause slight variations between traces measured from different boards. This difference is relevant when matching simulated traces with measured traces. However, we found no reference which evaluates the differences in traces between different sets of traces from different boards.  

\noindent    
\textbf{\textit{Challenge: Many of the measures provide evidence based on visual comparison. }} As the number of samples between the simulator output and the reference traces is different, it is difficult for the two sets of data to be compared directly.  Here we give some examples: 
\noindent
\begin{itemize}
\item \textit{Dynamic Time Warping.} To evaluate SILK~\cite{SILK_2014}, the distance between a set of measured traces and the simulator's output is computed. Dynamic Time Warping (DTW),  computes the distance between two temporal series, even when they have different elements. Several leakage models are used to generate simulated traces. DTW is instantiated with two different distance metrics, Euclidean and correlation-based distance. For a fair comparison, noise is added to the simulated traces. Notably, the two used distance metrics give different results regarding what constitutes a more realistic trace set.
\item \textit{Power correlation.}  Although the term is defined in~\cite{ACA_2020}, this is a well-known measure used when analyzing side-channel leakage. Here, it assumes the knowledge of the key and requires the explicit choice of a target variable and leakage model. The authors of ELMO~\cite{ELMO_2017} compare (visually) the correlation traces produced by predicting the ELMO leakage model on the measured traces with the same model's prediction on the traces produced by the simulator. The approach used then is to apply the same measure to both sets and show how the trend matches.  
\item \textit{Leakage detection comparison.} The de facto technique for leakage detection is TVLA~\cite{Goodwill_TVLA}. It comes then as no surprise that the practise of comparing a $t$-test trace produced by a leakage simulator with the $t$-test trace produced by the reference traces, using either a fixed-vs random test~\cite{ELMO_2017,ROSITA_2019} or a fixed-vs-fixed test~\cite{Reparaz_2016} is widespread. For computing the $t$-test traces, the datasets for both the simulator and the reference traces are prepared in advance by feeding the cryptographic algorithm with a fixed or a random plaintext. Next to its simplicity, this test is non-specific,  meaning that it does not target one specific variable. The classical application is a visual check to ensure that the simulated and reference $t$-traces match the identified leaking points.
\item   \textit{DPA performance.} To demonstrate the merit of a profiled simulator, Debande et al.~\cite{Debande_2012} compares the evolution of a DPA attack,  in terms of guessing entropy,  between a set of measured traces, a set of traces generated by a profiled leakage model,  and a set of traces generated by a non-profiled model.  Although the attack performance of the non-profiled leakage model is superior to the  that of the measured traces (and to that of the traces produced by the profiled leakage models),  the conclusion is that the profiled models which closely follow the behaviour of the measured traces are preferred. The main indicator for a desirable output is to match the trend of guessing entropy and the simulated traces' success rate with one of the real traces. We note that DPA is capable of tracking the performance for one target intermediate value. The same metric is used for evaluating the performance of NCSIM~\cite{NCSIM_2007} where the similarity of a DPA attack performed on the internal MOV operation is used to demonstrate the advantage of using a simulator for the design of a secure chip.  The authors of PARAM~\cite{PARAM_2020} also use DPA to compare the reference architecture's resistance before and after applying hardening countermeasures.  
\end{itemize}

Aside the fact that visual comparison is a subjective measure,  typically  due to space limitations, we only see one example of how these match. It is fair to point out that quantifiable measures for assessing leaks are sparse and not widely accepted in the side-channel community.

\noindent
\textbf{\textit{Challenge:  No consensus for what is a good procedure in comparing simulated vs reference traces.}}  An open question is whether the leakage model should be included in the evaluation. While most of the measures mentioned above do include a leakage model,  the authors of EMSIM~\cite{EMSIM_2020} use  \textit{normalized cross-correlation} to show how well the simulated traces match the reference traces without relying on a specific leakage model. As it computes the average cross-correlation between individual clock cycles, this metric's output is a quantifiable measure,  EMSIM reports an impressive 94.1\% accuracy in simulating side-channel signals across all possible instruction combinations. The requirement to use this measure is to precisely identify the correct clock cycles, which requires knowledge of the design details.   Another open questions, is whether the existing measures are enough or new ones are needed and  some contributions propose new metrics for evaluating the two sets.   Although initially a measure of the side-channel created by a single instruction the \textit{Signal Available to Attacker},  (SAVAT)~\cite{SAVAT_2014}  is used in EMSIM~\cite{EMSIM_2020} to measure the similarity between the simulated and reference traces. For a set of six instructions, the pairwise score SAVAT is computed, and the results between the simulated and reference traces are compared and found to be very close.  SCRIPT~\cite{SCRIPT_2020} uses \textit{side-channel vulnerability} (SCV), which is the equivalent of SNR~\cite{SNR_Mangard_2004} at the pre-silicon stage,  as it requires a small number of traces to compute and differs from SNR,  according to its authors by a scaling factor.  RTL-PSC~\cite{RTL_PSC_2019} combines KL-divergence  with SNR to identify vulnerable design blocks,  while SLEAK~\cite{Sleak_2014} uses mutual information between the sensitive values processed by the algorithm and the value or state of a system component during the execution binary.

\subsection{Metrics for evaluating the quality of the leakage model} 
It is common for the post-silicon evaluation tools~\cite{ELMO_2017,Debande_2012,ROSITA_2019} which propose a complex leakage model to compare its performance with simpler or previously known leakage models. In pre-silicon simulators, we count in this category the metrics which quantify the leakage identified by the simulator, compared to an ideal case, where the target does not leak information. 

\noindent
\textbf{\textit{Achievement: Empirical evidence that gray-level leakage models are superior to black-box leakage model.}}
Although the statement above might seem naive,  the question if \textit{it is worth to invest time and effort in creating  sophisticated gray leakage models} is valid. 
 To prove the merit of the ELMO leakage model~\cite{ELMO_2017}, the authors use \textit{power correlation},  to compare the predictions of a simple leakage model  (Hamming weigh) with the prediction of leakage produced by the ELMO model. The comparison is made by computing the correlation traces produced by both leakage models on \textit{the same} reference traces.  The result,  Figure~4 in~\cite{ELMO_2017} shows that the peaks in the correlation trace generated by the ELMO model are more clearly defined compared to those produced by the simple model. Additionally, the ELMO leakage model generates more peaks. The conclusion drawn by the authors is that the simple leakage model captures only a portion of the true leakage and should not be relied upon when protecting sensitive data. The same metric is used by  $\text{ELMO}^*$~\cite{ROSITA_2019} to shows it superiority to ELMO~\cite{ELMO_2017} leakage model.  Debande et al.~\cite{Debande_2012} use guessing entropy to compare the performance of a simple black-box leakage model with a profiled leakage model.  

\textbf{\textit{Achievement: Metrics to quantify leakage of hardware components.}}
\textit{Side Channel Vulnerability Factor} (SVF)~\cite{SVF_2012} quantifies the correlation between attacker observation patterns and patterns in victim execution. The insight is that side-channel attacks rely on recognizing leaked execution patterns.  SVF quantifies the patterns in attackers' observations and measures the correlation with the victim's actual execution patterns and captures systems' vulnerability to side-channel attacks. SVF quantifies the overall 'leakiness' of a particular system but does not provide insight into the cause. . In  PARAM~\cite{PARAM_2020}, SVF is used to quantify the amount of leakage in the target processor's different components. 

\textit{Leakage Impact Factor (LIF)} is used by the authors of ACA~\cite{ACA_2020} to quantify the similarity of the activity profile of a single gate or cell to a high-level leakage model used by DPA. The LIF is further weighted by the relative power consumption of the cell in the overall design. The LIF directly quantifies the contribution of a single gate or cell to the side-channel leakage, and it is used in ACA to rank the gates of the design from leaky to least leaky.

\subsection{Case studies} 
\label{subsec:Case_studies}

In this section,  we explore the answers to the question: \textit{How effective are the existing tools at verifying and eliminating SCA vulnerabilities}? To answer this question, most case studies will showcase the tool's ability to verify leakage (find the cause for producing leakage) and mitigate the leakage (eliminate it manually or automatically).  The depth and breadth of the presented cases vary greatly between the different contributions. While some verification tools showcase a toy example~\cite{PINPAS_2003}, others explore a wide variety of scenarios. For a leakage verification tool, the case study will reproduce a known flaw, introduce one in an otherwise secure design, or seldom use the tool to find a new unknown vulnerability.  In the following, we present a representative selection of use cases. 

\subsubsection{ Software implementation of cryptographic algorithms}

To show its effectiveness,   the tool of Reparaz~\cite{Reparaz_2016} is used to test the security of six high-order implementations. The first is a "smoke test" where the aim is to reproduce the flaw found by~\cite{Reparaz_RP10} for the first-order masking scheme proposed by~\cite{Reparaz_BFGV12}. The simulator performs six fixed-vs-fixed TVLA tests and reports that five of the six tests show leakage. The same test is applied to the first-order secure table recomputation scheme proposed by~\cite{Reparaz_Cor14} which, as expected, "on the strength of the found evidence" is reported to be secure. Next, the tool is used to reproduce the second-order flaw spotted by~\cite{Reparaz_CPRR13} in the masking scheme proposed by~\cite{Reparaz_RP10}. The authors report that it takes the tool 5 seconds to find the flaw, including the time to spot the cause. Next, the tool is used to reproduce the third-order vulnerability spotted by~\cite{Reparaz_Cor14} in the technique used for refreshing the mask in the scheme proposed by~\cite{Reparaz_RP10}. The authors report that the flaw was found in less than one second. A more difficult case for the tool (200 million traces and eight hours of simulation) is to reproduce the observations from~\cite{Reparaz_RBN15} on higher-order implementations~\cite{Reparaz_BGN14}. 
The authors also report a new second-order flaw found in~\cite{Reparaz_SP_06}, which, once found, is easily proved. 
\subsubsection{ Hardware implementation of cryptographic algorithms}

KARNA~\cite{KARNA_2019} is put to the test of securing three open-source cores. The first is a bit-serial implementation of the Simon block cipher~\cite{Karna_Simon} \footnote{reference implementation, \url{https://opencores.org/projects/simon_core}}. The tool performs three iterations, using TVLA with 8000 inputs and removes the leaking gate. The netlist is synthesized at 28 nm cell size. The second is a PRESENT core\footnote{reference implementation \url{https://opencores.org/projects/present}}, a minimal design which after one iteration achieves side-channel resilience.  The third is optimizing the AES  core~\cite{Karna_TinyAES}\footnote{reference implementation \url{https://opencores.org/projects/tiny_aes}} after which a DPA attack is performed  after place and route stage,  and the design is shown not to leak information with 100\,000 traces. Compared with the unoptimized AES synthesized design, the result is shown to reveal the correct key byte at approx 2k traces.  Karna can achieve a user-specified security level in all three designs with no impact on the delay or the number of gates and a 20\% increase in the utilization area.   

RTL-PSC~\cite{RTL_PSC_2019} is evaluated on two AES designs based on Galois Field (GF)~\cite{RTL_PSC_16} and Look-up Table(LUT)~\cite{RTL_PSC_17}. The tool is used to identify the leaky modules in the design, using a combination of Kullback-Leibler divergence and success rate. 
The validation of the tool is done on both gate-level netlist simulation and FPGA simulation. The comparison is made by computing the Pearson correlation of the RTL simulations (produced by the tool) with the gate-level netlist's KL-divergence trace.

AMASIVE~\cite{AMASIVE_2013} is showcased against an unprotected hardware implementation of the PRESENT cipher~\cite{Amasive_2}, for the first and the last round. The tool identifies hypothesis functions for the HW and HD leakage model. To confirm the tool's attack vector, a CPA attack is mounted, and the key is recovered within 10\,000 traces. 

\subsubsection{ Hardening of non-cryptographic hardware}

PARAM~\cite{PARAM_2020} is used to produce a hardened implementation of a Shakti-C~\cite{PARAM_Shakti_processors} core. The approach used to secure the software AES implementation deviates from the classical application of countermeasures. The authors identify and remove the leakage from each hardware component of the microprocessor. The DPA results in terms number of traces vs correlation score are used to show the hardened microprocessor's resilience.

\subsubsection{ Combination of software implementations running on a physical target}

The case study for MAPS focuses on showcasing the design flow with a tool such as MAPS. An example is a naive implementation of SIMON~\cite{MAPS_SIMON_SPECK} protected with Trichina AND gate~\cite{MAPS_Trichina_AND_gate}, which aims to minimize the number of execution cycles. The authors simulate the implementation of this cipher with and without pipeline leakage. Using a random vs fixed t-test, it is shown that both instances leak information. In the next iteration, the leakage is due to the reuse of register registers. In this version, the remaining leakage comes from the two pipeline registers. After fixing the two pipeline registers' leakage, the t-test traces obtained from the simulated traces show no leaky points as a final step t-test is also performed on a set of reference traces measured from a physical implementation, which show a few remaining leaky points. 

SAVRASCA~\cite{SAVRASCA_2017} is used to find a flaw in the AES implementation used for version 4 of the DPA contest\footnote{\url{http://www.dpacontest.org/v4/rsm_doc.php}}. Using the tool, the authors noted that the simulated traces' size depended on the value manipulated by the microcontroller, even though the implementation was running in a constant number of cycles.  Analysing the implementation, the authors found that the number of register access depended on the manipulated value. As a response to this finding, the new version of the DPA contest v4.2 fixed the implementation and released a new set of traces. 

ASCOLD~\cite{ASCOLD_2017} is used to develop a hardened  1st-order, ISW-based~\cite{Ascold_18} S-box with a bit-sliced RECTANGLE implementation~\cite{MAPS_rectangle}.  The performance of the hardened implementation is investigated for two different security objectives. The first is an \textit{efficient} implementation where the registers are cleared on a need-to basis to avoid overwrite and remnant effects. The second is \textit{conservative} implementation, which adds to the efficient implementation of dummy instructions' insertion through register/memory clearing. The strengths of hardening were evaluated using non-specific TVLA.  

COCO uses the open-source IBEX core\footnote{reference implementation \url{https://github.com/lowRISC/ibex}}, part of the PULP platform~\cite{Coco_PulpPlatform} and the OpenTitan~\cite{Coco_OpenTitan} project. The main application of COCO is the verification of a masked software implementation running on hardware specified at gate-level netlist.  A considerable selection of masked circuits, which cover domain-oriented masking (DOM) AND gate~\cite{Coco_20}, Ishai-Sahai-Wagner (ISW) AND~\cite{Reparaz_RBN15}, Threshold implementation(TI) AND~\cite{Coco_30} and larger implementations DOM Keccak S-box~\cite{Coco_21, DBLP:journals/cryptologia/BertoniDPA14},  DOM AES S-box~\cite{Cascade_13} and the Trichina AND gate~\cite{MAPS_Trichina_AND_gate} are presented to demonstrate the effectiveness of the tool. The scenarios cover two case studies with intentionally injected vulnerabilites~\cite{Coco_20,Coco_21}. The implementations also cover second-order security~\cite{Coco_20,Coco_21} and third-order security~\cite{Coco_20} (for a complete overview see Table 3 in~\cite{COCO_2020}).
 To show that COCO's output leads to practical secure implementations, a sample of the verified netlist of IBEX cores and the DOM Keccak S-box~\cite{Coco_21} is mapped onto a Xilinx Spartan-6  FPGA. The design is then evaluated using TVLA and shown to not leak information at 100000 traces.

\subsection{Evaluating Usability} 

Next to security-related advantages, emulators also offer important \textit{usability} advantages for the implementation and testing of a (masked) cryptografic primitive. The use of a simulator encourages testing at different development stages, allowing the removal of vulnerabilties as early as possible in the development cycle.  For post-silicon, the cryptographic implementation can be tested at source- code level~\cite{Reparaz_2016}, assembly level~\cite{ELMO_2017} or compiled binary. For pre-silicon, the design can be tested at RTL, gate or transistor level. 

\begin{itemize}
    \item \textit{Ease of use and convenience.} Building a setup for side-channel measurement is costly as it requires time, equipment, and expertise for preparing the target. Furthermore, the implementation and testing of a cryptographic primitive requires advanced skills in cryptographic engineering. The simulator is easy to use and reduces the effort of  a task that is highly iterative and requires (manual) effort.\\
    \textit{Application: post-silicon.}
    \item \textit{Fast(er) Development Cycles.} The speed of simulating the power consumption of a design is increasing as we progress with the design stages.  As the complexity of the design increases, \cite{ROSITA_2019} mentions a 4.5--7 speed increase compared to real hardware. As an extreme example~\cite{CASCADE_2020}  mentiones that for a fully-unrolled AES (which exceeds the security budget of most embedded device), simmulation and analysis of one million traces, can be done in 4 hours on a 8-thread worskstation. At the same time we have ample evidence ({\cref{subsec:Case_studies}}) that leakage at early design stages does have a positive effect.  The flexibility in making design changes and the leakage assessment time depends on the design stage~\cite{RTL_PSC_2019}. In other words,  while it is relatively easy to make changes at RTL-level, only small changes are possible at the layout level, while at post-silicon level, no changes in the design can be made. 
    \textit{Application: pre and post-silicon.}
    \item \textit{Cost.} Faster development cycles, and assurance in the final product ultimately increase the time to market of the product, which save costs or give a significant competitive advantage. For the post-silicon development stage, the idea of performing side-channel evaluation without the need of a lab and a team of experts available for assistance will make side-channel evaluation more accessible.\\
    \textit{Application: pre and post-silicon.}
\end{itemize}

\section{Open problems in designing side-channel simulators }
\label{sec:open_questions}

\noindent
\textbf{\textit{Open problem: Fine grained leakage models for complex architectures, with no design information.}} For post-silicon simulators, the main challenge for developing sophisticated gray-box leakage models is the lack of micro-architectural details.  For an accurate simulator at the post-silicon stage, the techniques and methods available aim \emph{to find and add},  the elements which capture leaks to gain precision. Today,  we know how to model simple micro-architectural features, such instruction-dependent activities in different pipeline stages,   add support for sequence dependency (the power consumed by an instruction depends on the other instructions in the pipeline), or find hidden storage elements. However,  creating a model such as  ELMO~\cite{ELMO_2017}  is prohibitively effort-intensive, even for relatively simple processors (in-order, no cache).  Furthermore,  it is unclear how to capture micro-architecture events characteristic for more complex processors, e.g.  pipeline stalls,  misprediction or cache miss.

\noindent
\textbf{\textit{Open problem: EM simulation. }} The one EM simulator we have,  EMSIM~\cite{EMSIM_2020}, a trace simulator, was constructed for a relatively simple, custom-made Risc-V processor.  To build the simulator,  EM measurements of the physical device are required.  Based on the experimental results aimed to assess how the simulated signal degrades when key micro-architectural features are omitted, the authors conclude that it is not possible to build an EM simulator without access to micro-architectural information.  This means that we may only be able to build EM simulators for open-source hardware.  If the presence of a physical target is a must for constructing EM simulators, it could explain the fact that there are no EM simulators at pre-silicon stage. 

\noindent
\textbf{\textit{Open problem: Metrics for quantifying potential side-channel leaks at micro-architectural  level. }} While the importance of micro-architecture features on the security of masked implementation has been shown to be crucial~\cite{Lazy_engineer+14, DeMeyer+20}  most contributions focus on one platform and zoom in on the feature which leaks side-channel information on the studied platform.  Even if  we would know how to model every micro-architecture event,  there is no widely accepted measure in the side-channel community to quantify leaks. 

\noindent
\textbf{\textit{Open problem: Benchmark existing simulators. }}
While some simulators are being open-sourced,  the specific scenarios for which the tools are intended for, are  hard to compare.  An alternative,  is to agree on representative,  public data-sets and  could be used to evaluate the potential of a tool, even when the tool is not open-sourced.

\noindent
\textbf{\textit{Open problem: Lacking case studies for asymmetric cryptographic implementations.  }}
All case studies we encountered in the existing literature are focused on the implementation of symmetric algorithms. 

\noindent
\textbf{\textit{Open problem: Composability of pre-silicon simulators.  }} 
While power simulation techniques are known and used, the primary application is heat dissipation and battery life. The goal of power estimation applied for side-channel evaluation is to capture the instantaneous power consumption. One of the important requirements is to process large quantities of data-dependent simulations.  This category's main challenge is to \textit{find and remove} the design specifications that do not contribute to leaks, such as to gain speed.  Today we have SCA-aware design-tools  for every design stage.  It has also been proven that removing vulnerabilities as early as possible in the design stage,  is a sound engineering practice.  Although creating a pre-silicon trace simulator requires significant effort,   the existing tooling is fragmented and not reusable.

\ifAnon
\else
\section*{Acknowledgments}
This research was supported by
The Australian Research Council Discovery Early Career Award DE200101577 
   and Discovery Project DP210102670,
the Blavatnik ICRC at Tel-Aviv University,
a gift from Intel Corporation, and the National Science Foundation Grant 1931639.
\fi
\bibliographystyle{plainnat}
\bibliography{SoK,case_studies}

\begin{thebibliography}{86}
\providecommand{\natexlab}[1]{#1}
\providecommand{\url}[1]{\texttt{#1}}
\expandafter\ifx\csname urlstyle\endcsname\relax
  \providecommand{\doi}[1]{doi: #1}\else
  \providecommand{\doi}{doi: \begingroup \urlstyle{rm}\Url}\fi

\bibitem[{Aoki Laboratory}(2007)]{RTL_PSC_16}
{Aoki Laboratory}.
\newblock Galois field based {AES} {Verilog} design.
\newblock \url{http://www.aoki.ecei.tohoku.ac.jp/crypto/web/cores.html}, 2007.
\newblock (accessed: 15.03.2021).

\bibitem[Aysu et~al.(2014)Aysu, Gulcan, and Schaumont]{Karna_Simon}
Aydin Aysu, Ege Gulcan, and Patrick Schaumont.
\newblock {SIMON} says: Break area records of block ciphers on {FPGAs}.
\newblock \emph{{IEEE} Embed. Syst. Lett.}, 6\penalty0 (2):\penalty0 37--40,
  2014.

\bibitem[Azouaoui et~al.(2020)Azouaoui, Bellizia, Buhan, Debande, Duval,
  Giraud, Jaulmes, Koeune, Oswald, Standaert, and Whitnall]{Buhan+20}
Melissa Azouaoui, Davide Bellizia, Ileana Buhan, Nicolas Debande,
  S{\'{e}}bastien Duval, Christophe Giraud, {\'{E}}liane Jaulmes,
  Fran{\c{c}}ois Koeune, Elisabeth Oswald, Fran{\c{c}}ois{-}Xavier Standaert,
  and Carolyn Whitnall.
\newblock A systematic appraisal of side channel evaluation strategies.
\newblock In \emph{SSR}, pages 46--66, 2020.

\bibitem[Balasch et~al.(2012)Balasch, Faust, Gierlichs, and
  Verbauwhede]{Reparaz_BFGV12}
Josep Balasch, Sebastian Faust, Benedikt Gierlichs, and Ingrid Verbauwhede.
\newblock Theory and practice of a leakage resilient masking scheme.
\newblock In \emph{Asiacrypt}, pages 758--775, 2012.

\bibitem[Balasch et~al.(2015)Balasch, Gierlichs, Grosso, Reparaz, and
  Standaert]{Lazy_engineer+14}
Josep Balasch, Benedikt Gierlichs, Vincent Grosso, Oscar Reparaz, and
  Fran{\c{c}}ois-Xavier Standaert.
\newblock On the cost of lazy engineering for masked software implementations.
\newblock In \emph{CARDIS}, pages 64--81, 2015.

\bibitem[Barbosa et~al.(2021)Barbosa, Barthe, Bhargavan, Blanchet, Cremers,
  Liao, and Parno]{Barbosa_2021_Computer_aided_crypto}
Manuel Barbosa, Gilles Barthe, Karthikeyan Bhargavan, Bruno Blanchet, Cas
  Cremers, Kevin Liao, and Bryan Parno.
\newblock {SoK}: Computer-aided cryptography.
\newblock In \emph{IEEE SP}, pages 123--141, 2021.

\bibitem[Barthe et~al.(2021)Barthe, Gourjon, Gr{\'{e}}goire, Orlt, Paglialonga,
  and Porth]{Barthe_2021_Masking_in_fine_grained_leakage_models}
Gilles Barthe, Marc Gourjon, Benjamin Gr{\'{e}}goire, Maximilian Orlt, Clara
  Paglialonga, and Lars Porth.
\newblock Masking in fine-grained leakage models: Construction, implementation
  and verification.
\newblock \emph{{IACR} Trans. Cryptogr. Hardw. Embed. Syst.}, 2021\penalty0
  (2):\penalty0 189--228, 2021.

\bibitem[Beaulieu et~al.(2015)Beaulieu, Shors, Smith, Treatman-Clark, Weeks,
  and Wingers]{MAPS_SIMON_SPECK}
Ray Beaulieu, Douglas Shors, Jason Smith, Stefan Treatman-Clark, Bryan Weeks,
  and Louis Wingers.
\newblock The {SIMON} and {SPECK} lightweight block ciphers.
\newblock In \emph{DAC}, 2015.

\bibitem[Bellard(2005)]{QEMU}
Fabrice Bellard.
\newblock {QEMU}, a fast and portable dynamic translator.
\newblock In \emph{USENIX ATC}, 2005.

\bibitem[Bernstein(2005)]{Bernstein05}
Daniel~J. Bernstein.
\newblock Cache-timing attacks on {AES}, 2005.
\newblock Preprint available at \url{http://cr.yp.to/papers.html\#cachetiming}.

\bibitem[Bernstein and Schwabe(2008)]{Karna_TinyAES}
Daniel~J. Bernstein and Peter Schwabe.
\newblock New {AES} software speed records.
\newblock In \emph{Indocrypt}, pages 322--336, 2008.

\bibitem[Bertoni et~al.(2005)Bertoni, Zaccaria, Breveglieri, Monchiero, and
  Palermo]{BZB+05}
Guido Bertoni, Vittorio Zaccaria, Luca Breveglieri, Matteo Monchiero, and
  Gianluca Palermo.
\newblock {AES} power attack based on induced cache miss and countermeasure.
\newblock In \emph{ITCC}, pages 586--591, 2005.

\bibitem[Bertoni et~al.(2014)Bertoni, Daemen, Peeters, and
  Assche]{DBLP:journals/cryptologia/BertoniDPA14}
Guido Bertoni, Joan Daemen, Micha{\"{e}}l Peeters, and Gilles~Van Assche.
\newblock The making of {KECCAK}.
\newblock \emph{Cryptologia}, 38\penalty0 (1):\penalty0 26--60, 2014.

\bibitem[Bhasin et~al.(2020)Bhasin, Chattopadhyay, Heuser, Jap, Picek, and
  Shrivastwa]{Bhasin_Mind_the_portability}
Shivam Bhasin, Anupam Chattopadhyay, Annelie Heuser, Dirmanto Jap, Stjepan
  Picek, and Ritu~Ranjan Shrivastwa.
\newblock Mind the portability: A warriors guide through realistic profiled
  side-channel analysis.
\newblock In \emph{NDSS}, 2020.

\bibitem[Bilgin et~al.(2014)Bilgin, Gierlichs, Nikova, Nikov, and
  Rijmen]{Reparaz_BGN14}
Beg{\"u}l Bilgin, Benedikt Gierlichs, Svetla Nikova, Ventzislav Nikov, and
  Vincent Rijmen.
\newblock Higher-order threshold implementations.
\newblock In \emph{Asiacrypt}, pages 326--343, 2014.

\bibitem[Binkert et~al.(2011)Binkert, Beckmann, Black, Reinhardt, Saidi, Basu,
  Hestness, Hower, Krishna, Sardashti, Sen, Sewell, Shoaib, Vaish, Hill, and
  Wood]{GEM5}
Nathan Binkert, Bradford Beckmann, Gabriel Black, Steven~K. Reinhardt, Ali
  Saidi, Arkaprava Basu, Joel Hestness, Derek~R. Hower, Tushar Krishna, Somayeh
  Sardashti, Rathijit Sen, Korey Sewell, Muhammad Shoaib, Nilay Vaish, Mark~D.
  Hill, and David~A. Wood.
\newblock The {Gem5} simulator.
\newblock \emph{ACM SIGARCH Computer Architecture News}, 39\penalty0
  (2):\penalty0 1--7, August 2011.

\bibitem[Bogdanov et~al.(2007)Bogdanov, Knudsen, Leander, Paar, Poschmann,
  Robshaw, Seurin, and Vikkelsoe]{Amasive_2}
A.~Bogdanov, L.~R. Knudsen, G.~Leander, C.~Paar, A.~Poschmann, M.~J.~B.
  Robshaw, Y.~Seurin, and C.~Vikkelsoe.
\newblock {PRESENT}: An ultra-lightweight block cipher.
\newblock In \emph{CHES}, pages 450--466, 2007.

\bibitem[Brumley and Tuveri(2011)]{BrumleyT11}
Billy~Bob Brumley and Nicola Tuveri.
\newblock Remote timing attacks are still practical.
\newblock In \emph{{ESORICS}}, pages 355--371, 2011.

\bibitem[Callan et~al.(2014)Callan, Zajic, and Prvulovic]{SAVAT_2014}
Robert~Locke Callan, Alenka~G. Zajic, and Milos Prvulovic.
\newblock A practical methodology for measuring the side-channel signal
  available to the attacker for instruction-level events.
\newblock In \emph{{MICRO}}, pages 242--254, 2014.

\bibitem[Camurati et~al.(2018)Camurati, Poeplau, Muench, Hayes, and
  Francillon]{DBLP:conf/ccs/CamuratiPMHF18}
Giovanni Camurati, Sebastian Poeplau, Marius Muench, Tom Hayes, and
  Aur{\'{e}}lien Francillon.
\newblock Screaming channels: When electromagnetic side channels meet radio
  transceivers.
\newblock In \emph{CCS}, pages 163--177, 2018.

\bibitem[Chari et~al.(2002)Chari, Rao, and Rohatgi]{ChariRR02}
Suresh Chari, Josyula~R. Rao, and Pankaj Rohatgi.
\newblock Template attacks.
\newblock In \emph{CHES}, pages 13--28, 2002.

\bibitem[Coron(2014)]{Reparaz_Cor14}
Jean-S{\'e}bastien Coron.
\newblock Higher order masking of look-up tables.
\newblock In \emph{Eurocrypt}, pages 441--458, 2014.

\bibitem[Coron et~al.(2014)Coron, Prouff, Rivain, and Roche]{Reparaz_CPRR13}
Jean-S{\'e}bastien Coron, Emmanuel Prouff, Matthieu Rivain, and Thomas Roche.
\newblock Higher-order side channel security and mask refreshing.
\newblock In \emph{Fast Software Encryption}, pages 410--424, 2014.

\bibitem[Daemen and Rijmen(2020)]{AES_DaemenR20}
Joan Daemen and Vincent Rijmen.
\newblock \emph{The Design of {Rijndael} - The Advanced Encryption Standard
  (AES), Second Edition}.
\newblock Information Security and Cryptography. Springer, 2020.
\newblock ISBN 978-3-662-60768-8.

\bibitem[Debande et~al.(2012)Debande, Berthier, Bocktaels, and
  Le]{Debande_2012}
Nicolas Debande, Maël Berthier, Yves Bocktaels, and Thanh-Ha Le.
\newblock Profiled model based power simulator for side channel evaluation.
\newblock Cryptology ePrint Archive, Report 2012/703, 2012.
\newblock \url{https://eprint.iacr.org/2012/703}.

\bibitem[Demme et~al.(2012)Demme, Martin, Waksman, and Sethumadhavan]{SVF_2012}
John Demme, Robert Martin, Adam Waksman, and Simha Sethumadhavan.
\newblock Side-channel vulnerability factor: A metric for measuring information
  leakage.
\newblock In \emph{{ISCA}}, pages 106--117, 2012.

\bibitem[den Hartog et~al.(2003)den Hartog, Verschuren, de~Vink, de~Vos, and
  Wiersma]{PINPAS_2003}
Jerry den Hartog, Jan Verschuren, Erik~P. de~Vink, Jaap de~Vos, and W.~Wiersma.
\newblock {PINPAS:} a tool for power analysis of smartcards.
\newblock In \emph{{SEC}}, pages 453--457, 2003.

\bibitem[{ETH Zurich}()]{Coco_PulpPlatform}
{ETH Zurich}.
\newblock Pulp platform.
\newblock \url{https://pulp-platform.org/}.
\newblock (accessed: 15.03.2021).

\bibitem[F et~al.(2020)F, Ganesan, Bodduna, and Rebeiro]{PARAM_2020}
Muhammad Arsath~K. F, Vinod Ganesan, Rahul Bodduna, and Chester Rebeiro.
\newblock {PARAM:} a microprocessor hardened for power side-channel attack
  resistance.
\newblock In \emph{{HOST}}, pages 23--34, 2020.

\bibitem[Fadl et~al.(2016)Fadl, Abu-Elyazeed, Abdelhalim, Amer, and
  Madian]{NCSIM_2007}
Omnia~S. Fadl, Mohamed~F. Abu-Elyazeed, Mohamed~B. Abdelhalim, Hassanein~H.
  Amer, and Ahmed~H. Madian.
\newblock Accurate dynamic power estimation for cmos combinational logic
  circuits with real gate delay model.
\newblock \emph{Journal of Advanced Research}, 7\penalty0 (1):\penalty0 89--94,
  2016.

\bibitem[Gagnerot(2013)]{Gagnerot_2013}
Georges Gagnerot.
\newblock \emph{\'Etude des attaques et des contre-mesures associ\'ees sur
  composants embarqu\'es}.
\newblock PhD thesis, Universit\'e de Limoges, 2013.

\bibitem[Gajski and Kuhn(1983)]{GajskiK83}
Daniel Gajski and Robert~H. Kuhn.
\newblock New {VLSI} tools - guest editors' introduction.
\newblock \emph{Computer}, 16\penalty0 (12):\penalty0 11--14, 1983.

\bibitem[Gala et~al.(2016)Gala, Menon, Bodduna, Madhusudan, and
  Kamakoti]{PARAM_Shakti_processors}
Neel Gala, Arjun Menon, Rahul Bodduna, G.~S. Madhusudan, and V.~Kamakoti.
\newblock {SHAKTI} processors: An open-source hardware initiative.
\newblock In \emph{{VLSI} Design}, pages 7--8. {IEEE} Computer Society, 2016.

\bibitem[Gandolfi et~al.(2001)Gandolfi, Mourtel, and Olivier]{GMO01}
Karine Gandolfi, Christophe Mourtel, and Francis Olivier.
\newblock Electromagnetic analysis: Concrete results.
\newblock In \emph{CHES}, pages 251--261, 2001.

\bibitem[Ge et~al.(2018)Ge, Yarom, Cock, and Heiser]{GeYCH18}
Qian Ge, Yuval Yarom, David Cock, and Gernot Heiser.
\newblock A survey of microarchitectural timing attacks and countermeasures on
  contemporary hardware.
\newblock \emph{J. Cryptographic Engineering}, 8\penalty0 (1):\penalty0 1--27,
  2018.

\bibitem[Genkin et~al.(2014)Genkin, Shamir, and Tromer]{GenkinST14}
Daniel Genkin, Adi Shamir, and Eran Tromer.
\newblock {RSA} key extraction via low-bandwidth acoustic cryptanalysis.
\newblock In \emph{{CRYPTO}}, pages 444--461, 2014.

\bibitem[Ghoshal and De~Cnudde(2017)]{Cascade_13}
Ashrujit Ghoshal and Thomas De~Cnudde.
\newblock Several masked implementations of the {Boyar}-{Peralta} {AES}
  {S-Box}.
\newblock In \emph{Indocrypt}, pages 384--402, 2017.

\bibitem[Gigerl et~al.(2020)Gigerl, Hadzic, Primas, Mangard, and
  Bloem]{COCO_2020}
Barbara Gigerl, Vedad Hadzic, Robert Primas, Stefan Mangard, and Roderick
  Bloem.
\newblock Coco: Co-design and co-verification of masked software
  implementations on {CPUs}.
\newblock Cryptology ePrint Archive, Report 2020/1294, 2020.
\newblock \url{https://eprint.iacr.org/2020/1294}.

\bibitem[Goodwill et~al.(2018)Goodwill, Jun, and P.Rohatgi]{Goodwill_TVLA}
G.~Goodwill, J.J.B. Jun, and P.Rohatgi.
\newblock A testing methodology for side channel resistance validation.
\newblock \emph{NIST non-invasive attack testing workshop}, 2018.

\bibitem[{Grant agreement ID: 507270}(2004)]{SCARD}
{Grant agreement ID: 507270}.
\newblock Side channel analysis resistant design flow.
\newblock \url{https://cordis.europa.eu/project/id/507270}, 2004.
\newblock [Online; accessed 3-Apr-2021].

\bibitem[Gross et~al.(2016)Gross, Mangard, and Korak]{Coco_20}
Hannes Gross, Stefan Mangard, and Thomas Korak.
\newblock Domain-oriented masking: Compact masked hardware implementations with
  arbitrary protection order.
\newblock In \emph{TIS}, 2016.

\bibitem[Gro{\ss} et~al.(2017)Gro{\ss}, Schaffenrath, and Mangard]{Coco_21}
Hannes Gro{\ss}, David Schaffenrath, and Stefan Mangard.
\newblock Higher-order side-channel protected implementations of {KECCAK}.
\newblock In \emph{{DSD}}, pages 205--212, 2017.

\bibitem[He et~al.(2019)He, Park, Nahiyan, Vassilev, Jin, and
  Tehranipoor]{RTL_PSC_2019}
Miao~Tony He, Jungmin Park, Adib Nahiyan, Apostol Vassilev, Yier Jin, and
  Mark~Mohammad Tehranipoor.
\newblock {RTL-PSC:} automated power side-channel leakage assessment at
  register-transfer level.
\newblock In \emph{{VTS}}, pages 1--6, 2019.

\bibitem[Huss et~al.(2013)Huss, St{\"{o}}ttinger, and Zohner]{AMASIVE_2013}
Sorin~A. Huss, Marc St{\"{o}}ttinger, and Michael Zohner.
\newblock \emph{{AMASIVE:} An Adaptable and Modular Autonomous Side-Channel
  Vulnerability Evaluation Framework}, volume 8260 of \emph{Lecture Notes in
  Computer Science}, pages 151--165.
\newblock Springer, 2013.

\bibitem[Ishai et~al.(2003)Ishai, Sahai, and Wagner]{Ascold_18}
Yuval Ishai, Amit Sahai, and David Wagner.
\newblock Private circuits: Securing hardware against probing attacks.
\newblock In \emph{CRYPTO}, pages 463--481, 2003.

\bibitem[Kocher(1996)]{Koc96}
Paul~C. Kocher.
\newblock Timing attacks on implementations of {Diffie}-{Hellman}, {RSA},
  {DSS}, and other systems.
\newblock In \emph{CRYPTO}, pages 104--113, 1996.

\bibitem[Kocher et~al.(1999)Kocher, Jaffe, and Jun]{KocherJJ99}
Paul~C. Kocher, Joshua Jaffe, and Benjamin Jun.
\newblock Differential power analysis.
\newblock In \emph{CRYPTO}, pages 388--397, 1999.

\bibitem[Kr{\"a}mer et~al.(2013)Kr{\"a}mer, Nedospasov, Schl{\"o}sser, and
  Seifert]{KN+13}
Juliane Kr{\"a}mer, Dmitry Nedospasov, Alexander Schl{\"o}sser, and Jean-Pierre
  Seifert.
\newblock Differential photonic emission analysis.
\newblock In \emph{COSADE}, pages 1--16, 2013.

\bibitem[Le~Corre et~al.(2018)Le~Corre, Gro{\ss}sch{\"{a}}dl, and
  Dinu]{MAPS_2018}
Yann Le~Corre, Johann Gro{\ss}sch{\"{a}}dl, and Daniel Dinu.
\newblock Micro-architectural power simulator for leakage assessment of
  cryptographic software on {ARM} {Cortex-M3} processors.
\newblock In \emph{{COSADE}}, pages 82--98, 2018.

\bibitem[Lerman et~al.(2015)Lerman, Poussier, Bontempi, Markowitch, and
  Standaert]{lerman2015template}
Liran Lerman, Romain Poussier, Gianluca Bontempi, Olivier Markowitch, and
  Fran{\c{c}}ois-Xavier Standaert.
\newblock Template attacks vs.\ machine learning revisited (and the curse of
  dimensionality in side-channel analysis).
\newblock In \emph{COSADE}, pages 20--33, 2015.

\bibitem[Lou et~al.(2021)Lou, Zhang, Jiang, and Zhang]{abs-2103-14244}
Xiaoxuan Lou, Tianwei Zhang, Jun Jiang, and Yinqian Zhang.
\newblock A survey of microarchitectural side-channel vulnerabilities, attacks
  and defenses in cryptography.
\newblock \emph{CoRR}, abs/2103.14244, 2021.

\bibitem[{lowRISC contributors}()]{Coco_OpenTitan}
{lowRISC contributors}.
\newblock Open titan.
\newblock \url{https://opentitan.org/}.
\newblock (accessed: 15.03.2021).

\bibitem[Maghrebi et~al.(2016)Maghrebi, Portigliatti, and
  Prouff]{maghrebi2016breaking}
Houssem Maghrebi, Thibault Portigliatti, and Emmanuel Prouff.
\newblock Breaking cryptographic implementations using deep learning
  techniques.
\newblock In \emph{SPACE}, pages 3--26, 2016.

\bibitem[Mangard(2004)]{SNR_Mangard_2004}
Stefan Mangard.
\newblock Hardware countermeasures against {DPA} -- a statistical analysis of
  their effectiveness.
\newblock In \emph{{CT-RSA}}, pages 222--235, 2004.

\bibitem[McCann et~al.(2017)McCann, Oswald, and Whitnall]{ELMO_2017}
David McCann, Elisabeth Oswald, and Carolyn Whitnall.
\newblock Towards practical tools for side channel aware software engineering:
  `grey box' modelling for instruction leakages.
\newblock In \emph{{USENIX} Security Symposium}, pages 199--216, 2017.

\bibitem[Meyer et~al.(2020)Meyer, Mulder, and Tunstall]{DeMeyer+20}
Lauren~De Meyer, Elke~De Mulder, and Michael Tunstall.
\newblock On the effect of the (micro)architecture on the development of
  side-channel resistant software.
\newblock Cryptology ePrint Archive, Report 2020/1297, 2020.

\bibitem[Murdock et~al.(2020)Murdock, Oswald, Garcia, Van~Bulck, Gruss, and
  Piessens]{DBLP:journals/ieeesp/MurdockOGBPG20}
Kit Murdock, David~F. Oswald, Flavio~D. Garcia, Jo~Van~Bulck, Daniel Gruss, and
  Frank Piessens.
\newblock Plundervolt: Software-based fault injection attacks against {Intel}
  {SGX}.
\newblock In \emph{{IEEE} S\&P}, pages 1466--1482, 2020.

\bibitem[Nahiyan et~al.(2020)Nahiyan, Park, He, Iskander, Farahmandi, Forte,
  and Tehranipoor]{SCRIPT_2020}
Adib Nahiyan, Jungmin Park, Miao~Tony He, Yousef Iskander, Farimah Farahmandi,
  Domenic Forte, and Mark~Mohammad Tehranipoor.
\newblock {SCRIPT:} a {CAD} framework for power side-channel vulnerability
  assessment using information flow tracking and pattern generation.
\newblock \emph{{ACM} Trans. Design Autom. Electr. Syst.}, 25\penalty0
  (3):\penalty0 26:1--26:27, 2020.

\bibitem[Nikova et~al.(2006)Nikova, Rechberger, and Rijmen]{Coco_30}
Svetla Nikova, Christian Rechberger, and Vincent Rijmen.
\newblock Threshold implementations against side-channel attacks and glitches.
\newblock In \emph{Information and Communications Security}, pages 529--545,
  2006.

\bibitem[Page(2002)]{Pag02}
Dan Page.
\newblock Theoretical use of cache memory as a cryptanalytic side-channel.
\newblock Cryptology ePrint Archive, Report 2002/169, 2002.
\newblock \url{http://eprint.iacr.org/2002/169/}.

\bibitem[Papagiannopoulos and Veshchikov(2017)]{ASCOLD_2017}
Kostas Papagiannopoulos and Nikita Veshchikov.
\newblock Mind the gap: Towards secure 1st-order masking in software.
\newblock In \emph{{COSADE}}, pages 282--297, 2017.

\bibitem[Quisquater and Samyde(2001)]{QS01}
Jean-Jacques Quisquater and David Samyde.
\newblock Electromagnetic analysis ({EMA}): Measures and counter-measures for
  smart cards.
\newblock In \emph{Smart Card Programming and Security}, pages 200--210, 2001.

\bibitem[Reparaz(2016)]{Reparaz_2016}
Oscar Reparaz.
\newblock Detecting flawed masking schemes with leakage detection tests.
\newblock In \emph{{FSE}}, pages 204--222, 2016.

\bibitem[Reparaz et~al.(2015)Reparaz, Bilgin, Nikova, Gierlichs, and
  Verbauwhede]{Reparaz_RBN15}
Oscar Reparaz, Beg{\"u}l Bilgin, Svetla Nikova, Benedikt Gierlichs, and Ingrid
  Verbauwhede.
\newblock Consolidating masking schemes.
\newblock In \emph{CRYPTO}, pages 764--783, 2015.

\bibitem[Rioja et~al.(2020)Rioja, Batina, and
  Armendariz]{Unai_When_similarities_are_taken_for_granted}
Unai Rioja, Lejla Batina, and Igor Armendariz.
\newblock When similarities among devices are taken for granted: Another look
  at portability.
\newblock In \emph{Africacrypt}, pages 337--357, 2020.

\bibitem[Riscure(2002)]{Inspector_SCA}
Riscure.
\newblock Inspector sca.
\newblock \url{https://www.riscure.com/security-tools/inspector-sca}, 2002.
\newblock [Online; accessed 7-Apr-2021].

\bibitem[Rivain and Prouff(2010)]{Reparaz_RP10}
Matthieu Rivain and Emmanuel Prouff.
\newblock Provably secure higher-order masking of {AES}.
\newblock In \emph{CHES}, pages 413--427, 2010.

\bibitem[Roth and Rudolph(2001)]{SimulAVR}
Theodore Roth and Klaus Rudolph.
\newblock {SimulAVR}.
\newblock \url{https://www.nongnu.org/simulavr/}, 2001.
\newblock [Online; accessed 13-Apr-2021].

\bibitem[{Satoh Lab}(2017)]{RTL_PSC_17}
{Satoh Lab}.
\newblock Lookup table based {AES} {Verilog} design.
\newblock \url{http://satoh.cs.uec.ac.jp/SAKURA/hardware/SAKURA-G.html}, 2017.
\newblock (accessed: 17.03.2021).

\bibitem[Schindler et~al.(2005)Schindler, Lemke, and Paar]{SLP05}
Werner Schindler, Kerstin Lemke, and Christof Paar.
\newblock A stochastic model for differential side channel cryptanalysis.
\newblock In \emph{CHES 2005}, pages 30--46, 2005.

\bibitem[Schramm and Paar(2006)]{Reparaz_SP_06}
Kai Schramm and Christof Paar.
\newblock Higher order masking of the {AES}.
\newblock In \emph{CT-RSA}, pages 208--225, 2006.

\bibitem[Sehatbakhsh et~al.(2020)Sehatbakhsh, Yilmaz, Zajic, and
  Prvulovic]{EMSIM_2020}
Nader Sehatbakhsh, Baki~Berkay Yilmaz, Alenka~G. Zajic, and Milos Prvulovic.
\newblock {EMSim}: A microarchitecture-level simulation tool for modeling
  electromagnetic side-channel signals.
\newblock In \emph{{HPCA}}, pages 71--85, 2020.

\bibitem[Shelton et~al.(2021)Shelton, Samwel, Batina, Regazzoni, Wagner, and
  Yarom]{ROSITA_2019}
Madura~A. Shelton, Niels Samwel, Lejla Batina, Francesco Regazzoni, Markus
  Wagner, and Yuval Yarom.
\newblock Rosita: Towards automatic elimination of power-analysis leakage in
  ciphers.
\newblock In \emph{NDSS}, 2021.

\bibitem[Sijacic et~al.(2020)Sijacic, Balasch, Yang, Ghosh, and
  Verbauwhede]{CASCADE_2020}
Danilo Sijacic, Josep Balasch, Bohan Yang, Santosh Ghosh, and Ingrid
  Verbauwhede.
\newblock Towards efficient and automated side-channel evaluations at design
  time.
\newblock \emph{J. Cryptogr. Eng.}, 10\penalty0 (4):\penalty0 305--319, 2020.

\bibitem[SLPSK et~al.(2019)SLPSK, Vairam, Rebeiro, and Kamakoti]{KARNA_2019}
Patanjali SLPSK, Prasanna~Karthik Vairam, Chester Rebeiro, and V.~Kamakoti.
\newblock Karna: A gate-sizing based security aware {EDA} flow for improved
  power side-channel attack protection.
\newblock In \emph{{ICCAD}}, pages 1--8, 2019.

\bibitem[Thuillet et~al.(2009)Thuillet, Andouard, and Ly]{OSCAR_2009}
C{\'{e}}line Thuillet, Philippe Andouard, and Olivier Ly.
\newblock A smart card power analysis simulator.
\newblock In \emph{{CSE} {(2)}}, pages 847--852, 2009.

\bibitem[Tiri et~al.(2005)Tiri, Hwang, Hodjat, Lai, Yang, Schaumont, and
  Verbauwhede]{DBLP:conf/ches/TiriHHLYSV05}
Kris Tiri, David~D. Hwang, Alireza Hodjat, Bo{-}Cheng Lai, Shenglin Yang,
  Patrick Schaumont, and Ingrid Verbauwhede.
\newblock Prototype {IC} with {WDDL} and differential routing - {DPA}
  resistance assessment.
\newblock In \emph{CHES}, pages 354--365, 2005.

\bibitem[Tiwari et~al.(1996)Tiwari, Malik, Wolfe, and
  Lee]{Tiwari_Instruction_level_power_analysis}
Vivek Tiwari, Sharad Malik, Andrew Wolfe, and Mike~Tien{-}Chien Lee.
\newblock Instruction level power analysis and optimization of software.
\newblock In \emph{{VLSI} Design}, pages 326--328, 1996.

\bibitem[Trichina et~al.(2005)Trichina, Korkishko, and
  Lee]{MAPS_Trichina_AND_gate}
Elena Trichina, Tymur Korkishko, and Kyung~Hee Lee.
\newblock Small size, low power, side channel-immune {AES} coprocessor: Design
  and synthesis results.
\newblock In \emph{Advanced Encryption Standard -- AES}, pages 113--127, 2005.

\bibitem[Veshchikov(2014)]{SILK_2014}
Nikita Veshchikov.
\newblock {SILK:} high level of abstraction leakage simulator for side channel
  analysis.
\newblock In \emph{PPREW@ACSAC}, pages 3:1--3:11, 2014.

\bibitem[Veshchikov and Guilley(2017)]{SAVRASCA_2017}
Nikita Veshchikov and Sylvain Guilley.
\newblock Use of simulators for side-channel analysis.
\newblock In \emph{EuroS{\&}P Workshops}, pages 104--112, 2017.

\bibitem[Walters et~al.(2014)Walters, Hagen, and Kedaigle]{Sleak_2014}
Dan Walters, Andrew Hagen, and Eric Kedaigle.
\newblock {SLEAK}: A side-channel leakage evaluator and analysis kit.
\newblock Technical report, The MITRE Corporation, November 2014.
\newblock [Online; accessed 8-Apr-2021].

\bibitem[Wegener and Moradi(2018)]{Cascade_39}
Felix Wegener and Amir Moradi.
\newblock A first-order {SCA} resistant {AES} without fresh randomness.
\newblock In \emph{COSADE}, pages 245--262, 2018.

\bibitem[Weissbart et~al.(2020)Weissbart, Chmielewski, Picek, and
  Batina]{WeissbartCPB20}
Leo Weissbart, Lukasz Chmielewski, Stjepan Picek, and Lejla Batina.
\newblock Systematic side-channel analysis of {Curve25519} with machine
  learning.
\newblock \emph{J. Hardw. Syst. Secur.}, 4\penalty0 (4):\penalty0 314--328,
  2020.

\bibitem[Yao et~al.(2020)Yao, Kathuria, Ege, and Schaumont]{ACA_2020}
Yuan Yao, Tarun Kathuria, Baris Ege, and Patrick Schaumont.
\newblock Architecture correlation analysis {(ACA):} identifying the source of
  side-channel leakage at gate-level.
\newblock In \emph{{HOST}}, pages 188--196, 2020.

\bibitem[Zhang et~al.(2015)Zhang, Bao, Lin, Rijmen, Yang, and
  Verbauwhede]{MAPS_rectangle}
Wentao Zhang, Zhenzhen Bao, Dongdai Lin, Vincent Rijmen, Bohan Yang, and Ingrid
  Verbauwhede.
\newblock {RECTANGLE:} a bit-slice lightweight block cipher suitable for
  multiple platforms.
\newblock \emph{Sci. China Inf. Sci.}, 58\penalty0 (12):\penalty0 1--15, 2015.

\end{thebibliography}

\end{document}